\documentclass[onefignum,onetabnum]{siamonline190516}

\pdfoutput=1

\usepackage{lipsum}
\usepackage{amsfonts}
\usepackage{graphicx}
\usepackage{epstopdf}
\usepackage{algorithmic}
\ifpdf
  \DeclareGraphicsExtensions{.eps,.pdf,.png,.jpg}
\else
  \DeclareGraphicsExtensions{.eps}
\fi

\usepackage[font={footnotesize,it}]{caption}
\usepackage{subcaption}

\usepackage{xr}
\usepackage{pdfpages}

\usepackage{enumitem}
\setlist[enumerate]{leftmargin=.5in}
\setlist[itemize]{leftmargin=.5in}

\newsiamremark{remark}{Remark}
\newsiamremark{hypothesis}{Hypothesis}
\crefname{hypothesis}{Hypothesis}{Hypotheses}
\newsiamthm{claim}{Claim}

\headers{Uncertainty quantification for wide-bin unfolding}{M. Stanley, P. Patil, and M. Kuusela}

\title{Uncertainty quantification for wide-bin unfolding: one-at-a-time strict bounds and prior-optimized confidence intervals\thanks{This work was supported by NSF grants DMS-2053804 and PHY-2020295, as well as JPL RSA No.\ 1670375.}}

\author{
Michael Stanley\thanks{Department of Statistics and Data Science, Carnegie Mellon University, Pittsburgh, PA 15213 \newline (\email{mcstanle@andrew.cmu.edu}).}
\and Pratik Patil\thanks{Department of Statistics and Data Science and Machine Learning Department, Carnegie Mellon University, Pittsburgh, PA 15213 (\email{pratik@cmu.edu}).}
\and Mikael Kuusela\thanks{Department of Statistics and Data Science and NSF AI Planning Institute for Data-Driven Discovery in Physics, Carnegie Mellon University, Pittsburgh, PA 15213 (\email{mkuusela@andrew.cmu.edu}).}
}

\usepackage{amsopn}

\ifpdf
\hypersetup{
  pdftitle={Uncertainty quantification for wide-bin unfolding: one-at-a-time strict bounds and prior-optimized confidence intervals},
  pdfauthor={Michael Stanley, Pratik Patil, and Mikael Kuusela}
}
\fi

\usepackage{amssymb}
\usepackage{latexsym}

\usepackage{amsmath}
\usepackage{amsfonts}
\usepackage{bm}
\usepackage{bbm} 
\usepackage{mathtools}
\usepackage{graphicx}
\usepackage{stackengine}

\usepackage{pifont}
\newcommand{\cmark}{\ding{51}}%
\newcommand{\xmark}{\ding{55}}
\renewcommand{\checkmark}{\cmark} 

\newcommand\barbelow[1]{\stackunder[1.2pt]{$#1$}{\rule{.8ex}{.075ex}}}

\renewcommand{\hat}{\widehat} 
\renewcommand{\tilde}{\widetilde} 
\renewcommand{\bar}{\overline} 
\renewcommand{\barbelow}{\underline} 
\let\eps\epsilon 
\renewcommand{\epsilon}{\varepsilon} 

\newcommand{\bepsilon}{\boldsymbol{\epsilon}}

\newcommand{\blambda}{\boldsymbol{\lambda}}
\newcommand{\bmu}{\boldsymbol{\mu}}

\newcommand{\bSigma}{\boldsymbol{\Sigma}}

\begin{document}

\maketitle

\begin{abstract}
Unfolding is an ill-posed inverse problem in particle physics aiming to infer a true particle-level spectrum from smeared detector-level data. For computational and practical reasons, these spaces are typically discretized using histograms, and the smearing is modeled through a response matrix corresponding to a discretized smearing kernel of the particle detector. This response matrix depends on the unknown shape of the true spectrum, leading to a fundamental systematic uncertainty in the unfolding problem. To handle the ill-posed nature of the problem, common approaches regularize the problem either directly via methods such as Tikhonov regularization, or implicitly by using wide-bins in the true space that match the resolution of the detector. Unfortunately, both of these methods lead to a non-trivial bias in the unfolded estimator, thereby hampering frequentist coverage guarantees for confidence intervals constructed from these methods. We propose two new approaches to addressing the bias in the wide-bin setting through methods called One-at-a-time Strict Bounds (OSB) and Prior-Optimized (PO) intervals. The OSB intervals are a bin-wise modification of an existing guaranteed-coverage procedure, while the PO intervals are based on a decision-theoretic view of the problem. Importantly, both approaches provide well-calibrated frequentist confidence intervals even in constrained and rank-deficient settings. These methods are built upon a more general answer to the wide-bin bias problem, involving unfolding with fine bins first, followed by constructing confidence intervals for linear functionals of the fine-bin counts. We test and compare these methods to other available methodologies in a wide-bin deconvolution example and a realistic particle physics simulation of unfolding a steeply falling particle spectrum.
\end{abstract}

\begin{keywords}
  Statistical Decision Theory, Ill-Posed Inverse Problems, Constrained Inference, Convex Programming, Frequentist Coverage
\end{keywords}

\section{Introduction} \label{sec:introduction}

Experimental high-energy physics studies the interactions and properties of fundamental particles. This is done by observing particle collision events using massive particle detectors, such as those at the Large Hadron Collider (LHC) at CERN. In these experiments, it is often of interest to measure functions, called differential cross sections, that represent the probability of producing particles with a certain energy, momentum, angle or other kinematic properties. Unfortunately, a particle detector can only produce noisy measurements of these kinematic properties. As a result, the function that is directly observable in the detector is a ``smeared'' or ``blurred'' version of the physical function of interest. The process of using observations from this smeared function and our knowledge of the detector response to infer the actual physical function of interest is called \emph{unfolding} \cite{Prosper2011,Cowan1998,Blobel2013,Zech2016}, which is well-recognized to be an ill-posed inverse problem~\cite{Kaipio2005,Engl2000}.
 
Let $f \in \mathcal{F}$ be the unknown true particle-level function of interest among a class of possible particle-level functions $\mathcal{F}$ and $g \in \mathcal{G}$ be the smeared detector-level function where $\mathcal{G}$ is a class of possible smeared functions. Then we can represent the relationship between $f$ and $g$ by $g = Kf$, where $K : \mathcal{F} \to \mathcal{G}$ is a linear operator that represents the smearing in the detector. In the simplest case, $K$ might be a convolution operator, but it can also have more complex structure. In most cases in high-energy physics, the functions $f$ and $g$ are discretized using histograms. This leads to a discretized version of the problem that we can represent as $\bm{\mu} = \bm{K}\bm{\lambda}$, where $\bm{\lambda} \in \mathbb{R}^n$ and $\bm{\mu} \in \mathbb{R}^m$ denote vectors of bin counts in the particle-level and detector-level histograms, respectively, and the elements of the response matrix $\bm{K} \in \mathbb{R}^{m \times n}$ represent the bin-to-bin smearing probabilities (see, e.g., Chapter~11 in \cite{Cowan1998}).

The most common approach to unfolding is to use a large number $(n)$ of particle-level histogram bins. This makes the problem severely ill-posed and one needs to use regularization to obtain physically plausible solutions. Commonly used techniques for regularized unfolding are two variants of Tikhonov regularization \cite{hocker_svd, schmitt_tunfold} (that perform explicit statistical regularization) and an expectation-maximization iteration with early stopping \cite{dagostini} (that performs a type of algorithmic regularization). Such regularization leads to a reduction in the variance of the unfolded estimators by introducing a bias in the estimation. This \emph{regularization bias} can be beneficial for point estimation, but it can lead to severely miscalibrated uncertainty quantification \cite{kuusela_panaretos,kuusela_stark,kuusela_phd_thesis}, with no easy workarounds (Section~\ref{sec:biases} gives more details).

An alternative approach, which is being used in an increasing number of LHC analyses (see, e.g., \cite{cms_2016,cms_2019}), is to instead discretize the problem using large particle-level bins, or equivalently, a small number $(n)$ of unfolded bins. This is motivated by the physical intuition that a detector with a certain resolution should not be able to resolve features smaller than its intrinsic resolution---thus it is futile to attempt to infer bins smaller than the detector's resolution. Mathematically, reducing the number of estimated parameters leads to implicit regularization of the problem and, as a result, the model can be inverted without the need for explicit regularization. Unfortunately, in the histogram discretization, the elements of the response matrix $\bm{K}$ depend on the unknown shape of the particle-level function $f$ within the particle-level bins, and the wider the particle-level bins, the stronger the dependence. In practice, one has to use an ansatz of $f$ to form $\bm{K}$ and the resulting estimator will have a \emph{wide-bin bias} stemming from the misspecification of this ansatz. Again, uncertainty quantification is severely hampered, as it is challenging to rigorously quantify the resulting systematic uncertainty.

In this work, we take a different approach to wide-bin unfolding. Instead of imposing external regularization (either through explicitly regularized estimators or implicitly through the structure of the response matrix as in previous wide-bin unfolding), we focus on inferring certain structured functionals of $\bm{\lambda}$ and let the geometry of the functional and the operator along with physical constraints on $\bm{\lambda}$ to self-regularize the ill-posed problem. The basic idea of our approach can be summarized as follows: We first discretize the problem using narrow particle-level bins. When these bins are small enough, the systematic uncertainty in $\bm{K}$ becomes negligible which eliminates the wide-bin bias. Then, we invert the forward model \emph{without explicit regularization}. Since $n$ is large, this gives a solution $\bm{\hat{\lambda}}$ with massive bin-wise fluctuations that tend to be anti-correlated across neighboring bins. Since the solution is not regularized, there is no regularization bias and hence the uncertainties are well-calibrated but difficult for humans to interpret and use due to their size. This solution can, however, be further used to constrain functionals $\theta = \theta(\bm{\lambda}) = \bm{h}^\top \bm{\lambda}$ of the narrow-bin particle-level histogram~$\bm{\lambda}$. When the functional is an aggregation, averaging or smoothing operation, the anti-correlated fluctuations in $\bm{\hat{\lambda}}$ will largely cancel out, leading to a well-constrained estimator $\hat{\theta} = \bm{h}^\top \bm{\hat{\lambda}}$ for $\theta$ with well-calibrated uncertainties. This basic idea was recently demonstrated in a remote sensing inverse problem in \cite{patil}; see also \cite{Takiya2004,Helene2007} for a related approach where $\bm{h}$ is a given low-pass filter. Of particular interest in unfolding are functionals $\theta$ that aggregate several neighboring small bins into large bins whose width is comparable to the detector resolution. From physical intuition and due to the aforementioned cancellations, it should be possible to infer these functionals with well-constrained uncertainties, even though the uncertainties of the individual narrow bins are huge. This leads to a principled solution of the unfolding problem that provides well-calibrated and well-constrained uncertainties which do not suffer from either the regularization bias of explicitly regularized unfolding or the wide-bin bias of previous wide-bin unfolding approaches.

In order to obtain a practically useful implementation of this approach, one needs to address two important methodological challenges. First, there are known physical constraints for the function $f$ which should ideally be taken into account to help regularize the problem. These constraints depend upon the particular unfolding problem, but one universal constraint is the non-negativity of $\bm{\lambda}$. Second, in order to diminish the wide-bin systematic uncertainty in $\bm{K}$, it might be desirable to use more particle-level bins than detector-level bins so that $n > m$. In this situation, $\bm{K}$ cannot be inverted as it always has a non-trivial null space leading to non-identifiability in inferring $\bm{\lambda}$. The approach described in \cite{patil, rust_oleary, rust_burrus} is designed to handle both of these complications. Briefly, the approach is based on using constrained optimization to directly construct confidence intervals for $\theta$ in a way that allows one to handle the null space of $\bm{K}$ and constraints on $\bm{\lambda}$. We call these resulting uncertainties One-at-a-time Strict Bounds (OSB) since they provide uncertainty calibration for one functional at a time. In the previous work \cite{patil}, this approach was demonstrated in a mildly rank-deficient situation with $m > n$ and a one-dimensional null space in the context of atmospheric remote sensing. In this paper, we further demonstrate that the approach works well even when the null space is high-dimensional with $n \gg m$.

The One-at-a-time Strict Bounds are empirically well-calibrated, but a rigorous proof of their coverage has been elusive \cite{patil, rust_oleary, tenorio}. As a novel alternative, we introduce in this paper a method that has provably correct coverage while addressing the aforementioned complications with the constraints on $\blambda$ and the null space of $\bm{K}$. The method is based on taking a decision-theoretic view of the problem. As such, it falls under the emerging area of Decision-Theoretic Uncertainty Quantification (DTUQ; see, e.g., \cite{bajgiran2021uncertainty}). In this method, we regard the confidence interval for $\theta$ as a decision rule and optimize its expected width with respect to a prior distribution on $\blambda$ \emph{among those rules that guarantee specified frequentist coverage}. We call the resulting uncertainty bounds Prior-Optimized (PO) confidence intervals. Even though this method uses a prior distribution to optimize the interval width, it is notably distinct from Bayesian uncertainty quantification. The decision-theoretic method guarantees frequentist coverage as we only consider rules that guarantee specified coverage, which is not the case for Bayesian methods \cite{kuusela_panaretos,patil}. We demonstrate that in the wide-bin unfolding problem, these Prior-Optimized intervals are only slightly wider than the One-at-a-time Strict Bounds, they provide similar empirical coverage as the One-at-a-time Strict Bounds while theoretically guaranteeing correct coverage, and they display little sensitivity to the choice of the prior. The utility of this approach is not limited to unfolding---indeed, the Prior-Optimized intervals are potentially widely applicable in constrained and rank-deficient linear inverse problems and are therefore of independent interest beyond the application presented in this paper.

The rest of this paper is structured as follows. Section~\ref{sec:unfolding_problem} presents the high-energy physics unfolding problem, providing the scientific motivation for this work. Section~\ref{sec:interval_intro} provides a description of both the One-at-a-time Strict Bounds and Prior-Optimized confidence intervals, and how they are related. Additionally, this section contains descriptions of three other intervals against which we compare our proposed intervals. Section~\ref{sec:story_section} demonstrates through a simple deconvolution problem how traditional wide-bin unfolding can produce intervals with poor empirical coverage because of the wide-bin bias from an even slightly misspecified ansatz. Furthermore, we show how the proposed intervals can address these challenges and provide further simulation studies to evaluate their coverage and expected width across a range of configurations. Section~\ref{sec:particle_spectrum} applies these methods to a more realistic particle physics application, that of unfolding a steeply falling inclusive jet differential cross-section. We show the flexibility of the intervals and their constraints, as this application includes additional monotonicity and convexity constraints on the underlying intensity function. Finally, Section~\ref{sec:discussion_conclusion} provides further discussion and conclusions regarding the results in this paper and avenues for future work. The Python scripts used to produce the results in this paper are available at \url{https://github.com/mcstanle/unfolding_osb_po_uq}.

\section{The High-Energy Physics Unfolding Problem} \label{sec:unfolding_problem}

\subsection{Forward Model for Unfolding} \label{sec:forward_model}
For a detailed overview of the unfolding problem setup, we refer readers to \cite{kuusela_phd_thesis, kuusela_panaretos, kuusela_stark}. We follow the same notation here. Broadly, data in experimental high energy physics can be modeled as an indirectly observed Poisson point process. The unfolding problem is an inverse problem that arises in particle physics \emph{measurement analyses}. The aim of these types of analyses is to estimate the true (unknown) probability distribution of some variable of interest, e.g., energy, scattering angle, particle mass, or decay length~\cite{Blobel2013}. ``Folding" occurs when we observe one of these probability distributions through a detector where the observations are corrupted by stochastic noise. Then, unfolding is the process of inferring the true distribution that created the \emph{smeared} or \emph{blurred} observed distribution. This inverse problem is ill-posed because large changes in the true distribution may only result in small changes in the observed distribution \cite{kuusela_phd_thesis}.

Formally, this setup is described by a Poisson point process $M$, representing the true particle-level spectrum of events, and a Poisson point process $N$, representing the detector-level spectrum, which is related to $M$ via a smearing kernel $k$. More precisely, let $T \subseteq \mathbb{R}$, a compact interval, be the state space of $M$, and $S \subseteq \mathbb{R}$, a compact interval, be the state space of $N$. Each of these Poisson point processes is uniquely characterized by an intensity function; we denote by $f: T \rightarrow \mathbb{R}_+$ the intensity function for process $M$ and by $g : S \rightarrow \mathbb{R}_+$ the intensity function for process $N$. These intensity functions are the Radon--Nikodym derivatives of the mean measures $\lambda$ and $\mu$ of each point process, respectively.
For instance, for some event $B \in \sigma(T)$ (the Borel $\sigma$-algebra over $T$), we have $\lambda(B) = \int_B f(t) \, dt$. As explained in Section 3.1 of \cite{kuusela_phd_thesis}, letting variable $X$ to be a true particle-level observation, $X$ has the probability density \smash{$p_{X}(t) = f(t) / \lambda(T)$},
and hence we have that
\[
\mathbb{P}\left\{ X \in B \right\} = \int_B p_{X}(t) \, dt = \int_B f(t) \, dt / \lambda(T) = \lambda(B) / \lambda(T).
\]
As $X$ passes through the detector, it is subjected to noise that produces an observation $Y$ in the smeared space. Note, for simplicity of exposition, we assume that the detector always observes the true event and that no event is smeared over the boundaries of the smeared state space. Both of these effects can be rigorously treated using \textit{thinning} as described in detail in \cite{kuusela_masters_thesis, reiss}. The noise is assumed to be independent and identically distributed, and hence we obtain the following probability density for $Y$:
\begin{equation} \label{eq:smear_density}
    p_Y(s)
    = \int_T p_{X, Y}(t, s) \, dt 
    = \int_T p_{Y \mid X}(s \mid t) p_X(t) \, dt.
\end{equation}
Analogous to the relationship between the marginal distribution $p_X$ and the intensity function $f$, we have \smash{$p_Y(s) = g(s) / \mu(S)$}. With no thinning, since $Y$ is simply a perturbed version of $X$, $\lambda(T) = \mu(S)$, and hence we have $\lambda(T) p_Y(s) = g(s)$. Combining this equality with Eq.~\eqref{eq:smear_density}, we see that $g(s) = \int_T p_{Y \mid X}(s \mid t) f(t) \, dt$.
This shows that the true intensity $f$ and smeared intensity $g$ are connected via the conditional distribution of $Y$ given $X$. It is reasonable to assume that the detector injects Gaussian noise to each particle event, hence this conditional distribution can be safely assumed to be Gaussian. Furthermore, this Gaussian distribution defines the smearing kernel, $k : S \times T \rightarrow \mathbb{R}_+$, and connects the two intensity functions in the form of a Fredholm integral operator, $g(s) = \int_T k(s, t) f(t) \, dt$,
where $k(s, t) = p_{Y \mid X}(s \mid t)$. As such, observations from the process characterized by the intensity function $g(\cdot)$ are used in unfolding to infer the intensity function $f(\cdot)$ of the particle-level spectrum.

High-energy physics data are typically binned, discretizing the Poisson point processes. Let \smash{$\{ T_j \}_{j = 1}^n$} be a partition of the true space and \smash{$\{ S_i \}_{i = 1}^m$} be a partition of the smeared space. By the definition of Poisson point processes, we have for each $j \in [n]$ (where we use the convention $[n]$ to denote the set $\{ 1, \dots, n \}$ for a positive integer $n$) $M(T_j) \sim \textrm{Poisson}(\lambda(T_j))$,
where \smash{$\lambda(T_j) = \int_{T_j} f(t) \, dt$}. Additionally, since \smash{$\{ T_j \}_{j = 1}^n$} partitions $T$, $M(T_i)$ and $M(T_j)$ are independent for all $i \neq j$. Define the vector \smash{$\blambda = \left[\lambda(T_1), \dots, \lambda(T_n) \right]^\top$}. Likewise, let \smash{$\mu(S_i) = \int_{S_i} g(s) \, ds$} for $i \in [m]$ and define \smash{$\bmu = \left[\mu(S_1), \dots, \mu(S_m) \right]^\top$}. Thus, we can model particle-level and detector-level histograms, represented as random vectors $\boldsymbol{x}$ and $\boldsymbol{y}$ with elements $M(T_j)$, $j \in [n]$, and $N(S_i)$, $i \in [m]$, respectively, as follows:
 \begin{align}
     \boldsymbol{x} \sim \textrm{Poisson}(\blambda),
     \qquad
     \boldsymbol{y} \sim \textrm{Poisson}(\bmu). \label{eq:smear_draw}
 \end{align}
But, we also have
{\allowdisplaybreaks
\begin{align}
    \mu(S_i) &= \int_{S_i} g(s) ds = \int_{S_i} \int_T k(s, t) f(t) \, dt \, ds \\
    &= \int_{S_i} \left( \sum_{j = 1}^n \int_{T_j} k(s, t) f(t) \, dt \right) ds = \sum_{j = 1}^n \int_{S_i} \int_{T_j} k(s, t) f(t) \, dt \, ds \\
    &= \sum_{j = 1}^n \int_{S_i} \int_{T_j} k(s, t) f(t) \, dt \, ds \left( \int_{T_j} f(t) \, dt \right)^{-1} \lambda(T_j).
\end{align}}Hence, if we define
\begin{equation} \label{eq:smearing_matrix_element}
    K_{ij} = \frac{\int_{S_i} \int_{T_j} k(s, t) f(t) \, dt \, ds}{ \int_{T_j} f(t) \, dt}, \quad i \in [m], \quad j \in [n]
\end{equation}
a linear system of equations relates the two discretized Poisson point processes
\begin{equation} \label{eq:means_linear_system}
    \bmu = \bm{K} \blambda,
\end{equation}
where $\bm{K} \in \mathbb{R}^{m \times n}$ has the $ij$-th element given by Eq.~\eqref{eq:smearing_matrix_element}. Additionally, elements $K_{ij}$ have a probabilistic interpretation. Namely, the probability that an event in the true bin $T_j$ propagates to the smeared bin $S_i$ is given by $K_{ij}$ (see Proposition 2.11 in \cite{kuusela_masters_thesis}), i.e., we have $K_{ij} = \mathbb{P} \left( Y \in S_i \mid X \in T_j \right)$.
This discretization allows us to re-express Eq.~\eqref{eq:smear_draw} as
\begin{equation} \label{eq:true_poisson_data_gen}
    \mathbf{y} \sim \textrm{Poisson}\left( \bm{K} \blambda \right).
\end{equation}
This statistical model tells us that we observe Poisson distributed bin counts with mean~$\bm{K} \blambda$. Given this model, we wish to make inferences on the particle-level histogram~$\blambda$.

\subsubsection{Approximations}
To frame this statistical model in a more tractable form, we employ the Normal approximation to the Poisson distribution, which holds well in this application since each bin of observations typically contains a large number of events. Hence, we re-express the model as
\begin{equation} \label{eq:normal_stat_model}
    \mathbf{y} = \bm{K} \blambda + \bepsilon 
\end{equation}
\noindent where $\bepsilon \sim \mathcal{N}(\mathbf{0}, \bSigma)$ and $\bSigma = \textrm{diag}(\bm{K} \blambda)$ is an $m \times m$ diagonal matrix with $\bm{K} \blambda$ on the diagonal. We use the statistical model described Eq.~\eqref{eq:normal_stat_model} to perform inference on $\blambda$. For the simulations in later sections, we assume $\bm{K} \blambda$ is known, but this vector can easily be estimated via the observations $\mathbf{y}$ as shown in \cite{kuusela_phd_thesis}.

As described in \cite{kuusela_phd_thesis}, the response matrix $\bm{K}$ is typically obtained using detector simulations or other knowledge about the response behavior of the detector. But, as can be seen from Eq.~\eqref{eq:smearing_matrix_element}, $\bm{K}$ depends on the unknown intensity function $f$. There are several ways to get around this when computing $\bm{K}$, of which we describe the one that we use in this paper.

A Monte Carlo (MC) ansatz of the intensity function, denoted $f^{\textrm{MC}}$, can be used in the computation of $\bm{K}$. This is what happens when a Monte Carlo event generator is used to generate particle-level events which are then smeared using a detector simulator to obtain an estimate of $\bm{K}$. Hence, a different approximation to Eq.~\eqref{eq:smearing_matrix_element} is given by
\begin{equation} \label{eq:K_matrix_component}
    K_{ij} \approx \frac{\int_{S_i} \int_{T_j} k(s, t) f^{\textrm{MC}}(t) \, dt \, ds}{ \int_{T_j} f^{\textrm{MC}}(t) \, dt}, \quad i \in [m], \quad j \in [n].
\end{equation}
The matrix elements in Eq.~\eqref{eq:K_matrix_component} are usually obtained by tracking propagation of events across bins in a Monte Carlo simulation. We assume the simulation is large enough so that the Monte Carlo noise in these values is negligible. This approximation improves as the number of true bins ($n$) increases. Hence, we would like to use the finest possible (by computation or otherwise) discretization of the true space, leading to a rank-deficient $\bm{K}$~matrix.

\subsection{Regularization Bias, Wide-Bin Bias and Wide-Bins-via-Fine-Bins Unfolding} \label{sec:biases}

As described in the introduction, using a large number of true bins $(n)$ renders the problem severely ill-posed. For finite linear operators, this ill-posedness is the result of small values in the spectrum of singular values of $\bm{K}$ \cite{Kaipio2005,Engl2000,Hansen1998}, causing large fluctuations in the point estimators. Regularization methods like Tikhonov regularization essentially replace the small singular values with a term that is a function of the singular values and the regularization parameter, helping to tamp down the large fluctuations. Statistically, regularization introduces a bias to bring down the variance of the estimator. As illuminated in Section 4 of \cite{kuusela_phd_thesis}, Gaussian confidence intervals generated by these regularized estimators suffer from a loss of coverage as the regularization strength increases, due to the increasing \emph{regularization bias} of the estimator. More precisely, letting $\hat{\blambda}_j$ be the point estimator for the $j$-th bin count as estimated via a regularization method, Kuusela shows in Section 4 of \cite{kuusela_phd_thesis} that in most cases
\begin{equation}
    \mathbb{P} \Big( \blambda_j \in \big[\hat{\blambda}_j - z_{1 - \alpha / 2} \, \text{se}(\hat{\blambda}_j ), \, \hat{\blambda}_j + z_{1 - \alpha / 2} \, \text{se}(\hat{\blambda}_j ) \big] \Big) \ll 1 - \alpha,
\end{equation}
where $1 - \alpha$ is the desired confidence level for the interval, $z_{1-\alpha/2}$ is $(1 - \alpha/2)$ quantile of the standard Gaussian distribution and $\text{se}(\hat{\blambda}_j)$ is the standard error of $\hat{\blambda}_j$.
This happens because the estimator $\hat{\blambda}_j$ is inherently biased and it is very difficult to account for this regularization bias in order to construct regularized confidence intervals with accurate coverage.
Alternatively, one can reason that these small singular values requiring explicit regularization can be avoided by only inverting with a bin resolution commensurate with the resolution of the detector with which the data are being observed. Using these types of wide bins acts as implicit regularization. Specifically, instead of discretizing $\blambda \in \mathbb{R}^n$, we discretize $\tilde{\blambda} \in \mathbb{R}^k$, where $k \ll n$. Unfortunately, this approach is still exposed to bias since using wide bins makes $\bm{K}$ increasingly dependent on the assumed MC ansatz and hence the standard wide-bin approach can suffer from under-coverage in some bins due to this \emph{wide-bin bias}. Although, instead of the bias coming from explicit regularization as described above for Tikhonov regularization, for instance, it comes in this case from the potentially large systematic error in $\bm{K}$.

Our proposed general solution is to first invert, without regularization, using the same number of unfolded bins as in typical regularization-based inversion, but to then aggregate and propagate the errors of this first inversion to a bin resolution similar to the wide-bin strategy above. This strategy is implemented by defining a set of functionals aggregating groups of fine-resolution bins to the desired wide-bin resolution. More precisely, using the notation above, for each wide bin, $l \in [k]$, we define an associated functional $\theta_l(\blambda) = \bm{h}_l^\top \blambda$, where $\bm{h}_l \in \mathbb{R}^{n}$ is an aggregation weight vector and find a confidence interval for each such functional. A key feature of the two intervals presented herein is the direct optimization of the interval endpoints, as opposed to finding a point estimator and then subtracting and adding the appropriately scaled standard error of the estimator. This enables us to avoid explicitly performing the first inversion step which allows us to handle rank-deficient smearing matrices.

\section{Proposed Methods - One-at-a-time Strict Bounds and Prior-Optimized Intervals} 
\label{sec:interval_intro}
In Section~\ref{sec:story_section}, we motivate the need and utility of both the One-at-a-time Strict Bounds (OSB) and Prior-Optimized (PO) intervals through an expressive toy problem and least-squares intervals, which are the simplest analytically tractable confidence intervals one can construct (using the least-squares estimator). But these intervals pay for their simplicity in practice since they are unable to incorporate the null space of $\bm{K}$ and any constraints on the feasible space and their coverage guarantees break down under sufficient systematic error.

To address these concerns we explore the use of OSB intervals, described in \cite{patil, rust_oleary, rust_burrus, tenorio}, and expand on these intervals with PO intervals. The OSB intervals can leverage physical constraints and elegantly handle rank-deficient response matrices, the use of which provides a mitigation technique for handling systematic error. The intervals have correct empirical coverage; however, we are presently unable to theoretically prove their coverage. By contrast, we can prove coverage for the PO intervals even for constrained and rank-deficient situations. Empirical exploration has shown the PO intervals to be slightly wider than the OSB intervals, but with the advantage of coverage guarantees that are important for scientific applications.

Using optimization to directly find interval endpoints for statistical models shown in Eq.~\eqref{eq:normal_stat_model} is not a novel concept. Notably, Donoho provides a rigorous treatment of confidence intervals for linear functionals under model Eq.~\eqref{eq:normal_stat_model} in a minimax sense in \cite{donoho_94}. Stark develops the ``strict-bounds" approach in \cite{stark_strict_bounds} which develops techniques for intervals with simultaneous coverage. In exploring the utility of the OSB and PO intervals, we compare, when possible, with the minimax and strict-bounds intervals from these works. We describe the constructions of these intervals in Section \ref{sec:other_intervals}.

\subsection{One-at-a-time Strict-Bounds Intervals (OSB)} \label{sec:osb_description}
We adapt the nomenclature of \cite{patil} to fit the terms we have already defined. As a historical note, these intervals have also been described by Rust and O'Leary \cite{rust_oleary}, Rust and Burrus \cite{rust_burrus}, and Tenorio et al.\ \cite{tenorio} for a simpler version of the problem with non-negativity constraints. For ease of description, we consider a statistical model described per Eq.~\eqref{eq:normal_stat_model} for which the covariance matrix is the identity matrix $\mathbf{I}_m \in \mathbb{R}^{m \times m}$. This assumption is warranted since we can always whiten the observation vector $\mathbf{y}$. More precisely, consider the Cholesky decomposition of the covariance matrix: $\bSigma = \mathbf{L} \mathbf{L}^\top$ where $\bm{L} \in \mathbb{R}^{m \times m}$ is a lower triangular matrix, and consider
\begin{equation} \label{eq:cholesky_transform_data}
    \mathbf{L}^{-1} \mathbf{y} = \mathbf{L}^{-1} \bm{K} \blambda + \bm{\eta},
\end{equation}
where \smash{$\bm{\eta} \sim \mathcal{N}\big(\mathbf{0}, \mathbf{L}^{-1} \bSigma {(\mathbf{L}^{-1})}^\top\big)$}. Since $\mathbf{L}^{-1} \bSigma {(\mathbf{L}^{-1})}^\top = \mathbf{I}_m$, we have that $\bm{\eta} \sim \mathcal{N}( \mathbf{0}, \mathbf{I}_m)$. As such, when building upon the statistical model in Eq.~\eqref{eq:normal_stat_model}, we assume that $\bepsilon \sim \mathcal{N}(\mathbf{0}, \mathbf{I}_m)$.

For the OSB intervals, we wish to perform inference on a single functional of the true underlying parameter $\blambda$. We denote this functional by $\theta = \theta(\blambda)$ and parameterize it using the vector $\bm{h} \in \mathbb{R}^n$. This functional thus defines a quantity of interest, $\theta := \bm{h}^\top \blambda$, for which we wish to build a confidence interval $ [ \barbelow{\theta}, \bar{\theta} ]$.
Furthermore, the confidence interval should be constructed such that for a fixed $\alpha \in [0, 1]$ and any $\blambda$ such that $\mathbf{A} \blambda \leq \mathbf{b}$,
\begin{equation}
    \mathbb{P}_{\mathbf{y}} \left( \theta \in [\barbelow{\theta}, \bar{\theta} ] \right) \geq 1 - \alpha, \label{eq:correctCoverage}
\end{equation}
where $\mathbf{A} \in \mathbb{R}^{q \times n}$.
Here $\mathbf{A} \bm{\lambda} \le \bm{b}$ reflect any linear constraints that one has
on the parameter $\bm{\lambda}$. We elaborate more below.
To find these interval endpoints, \cite{patil} sets up two optimization problems, one to find the lower endpoint and one to find the upper endpoint. To find $\barbelow{\theta}_{\mathrm{OSB}}$, we solve the following optimization problem:
\begin{align} \label{opt:osb_lower}
    \textrm{minimize} \quad &\bm{h}^\top \blambda \nonumber \\
    \textrm{subject to} \quad &\lVert \mathbf{y} - \bm{K} \blambda \rVert_2^2 \leq z_{1 - \alpha / 2}^2 + s^2,  \\
    & \mathbf{A} \blambda \leq \mathbf{b}, \nonumber
\end{align}
where $z_{1 - \alpha / 2}$ is the quantile at level $(1 - \alpha / 2)$ of the standard Gaussian distribution, the matrix $\bm{K}$ is assumed to be transformed as in Eq.~\eqref{eq:cholesky_transform_data} corresponding to data with identity covariance, and $s^2$ is defined as the optimum value of the following optimization~problem:
\begin{align} \label{opt:slack}
    \textrm{minimize} \quad & \lVert \mathbf{y} - \bm{K} \blambda \rVert_2^2 \\
    \textrm{subject to} \quad & \mathbf{A} \blambda \leq \mathbf{b}. \nonumber
\end{align}
Similarly, to find $\bar{\theta}_{\mathrm{OSB}}$, we solve the following optimization problem:
\begin{align} \label{opt:osb_upper}
    \textrm{maximize} \quad &\bm{h}^\top \blambda \nonumber \\
    \textrm{subject to} \quad &\lVert \mathbf{y} - \bm{K} \blambda \rVert_2^2 \leq z_{1 - \alpha / 2}^2 + s^2,  \\
    & \mathbf{A} \blambda \leq \mathbf{b}. \nonumber
\end{align}
The linear inequality constraint $\mathbf{A} \blambda \leq \mathbf{b}$ allows the analyst to implement a wide range of restrictions on the feasible set. As we explore in Section~\ref{sec:particle_spectrum}, these constraints can enforce a variety of shape constraints on the vector $\blambda$, such as non-negativity, monotonicity, and convexity. For instance, to enforce non-negativity, which always holds true for the unfolding problem, we set the linear inequality constraint with $\mathbf{A} = - \mathbf{I}_n$ and $\mathbf{b} = \mathbf{0}$ so that the inequality constraint becomes $\blambda \geq \mathbf{0}$.
Both optimization problems \eqref{opt:osb_lower}, \eqref{opt:osb_upper},
and \eqref{opt:slack} are convex optimization problems,
and additionally can be cast as second-order cone programs.

Tenorio et al.\ \cite{tenorio} proved the coverage of these intervals under a stochastic ordering criterion which leads to the requirement that $\bm{h}$ be in the row space of $\bm{K}$. For scientists interested in a specific functional, this condition is restrictive when $\bm{K}$ does not have full column rank. Yet, both later in this paper and in \cite{patil}, it is observed empirically that these intervals in practice have the desired coverage even in rank-deficient situations. Indeed, in extensive testing, we have yet to find a situation where the coverage would breakdown. Unfortunately, rigorous proof in this case seems difficult, and hence we attest that these intervals do not yet have provable coverage in a wide range of scientifically-relevant situations.

Generalizing slightly, given $\xi \in \mathbb{R}_{+}$, define the set $\mathcal{Z}(\xi)$ such that
\begin{equation} \label{eq:interval_feasible_set}
    \mathcal{Z}(\xi) = \big\{ \blambda : \lVert \mathbf{y} - \bm{K} \blambda \rVert_2^2 \leq \xi \text{ and } \mathbf{A} \blambda \leq \mathbf{b} \big\}.
\end{equation}
Then, defining $\psi_{1 - \alpha/2}^2 := z^2_{1 - \alpha / 2} + s^2$, the OSB intervals can be compactly represented as follows:
\begin{equation} \label{eq:osb_construct}
    \big[ \barbelow{\theta}_{\mathrm{OSB}}, \bar{\theta}_{\mathrm{OSB}} \big]
    = \left[ \underset{\blambda \in \mathcal{Z}\big(\psi_{1 - \alpha/2}^2\big)}{\text{min}} \bm{h}^\top \blambda, \underset{\blambda \in \mathcal{Z}\big(\psi_{1 - \alpha/2}^2\big)}{\text{max}} \bm{h}^\top \blambda \right],
\end{equation}

We can also consider the Lagrangian dual programs for both optimization problems \eqref{opt:osb_lower} and \eqref{opt:osb_upper} as presented in \cite{patil}. To find $\barbelow{\theta}$, the dual problem is formulated as
\begin{align} \label{opt:dual_lower}
    \underset{\bm{w}, \bm{c}}{\text{maximize}} \quad & \bm{w}^\top \mathbf{y}
    - 
    \psi_{1 - \alpha / 2}
    \lVert \bm{w} \rVert_2 - \mathbf{b}^\top \bm{c} \nonumber \\
    \textrm{subject to} \quad & \bm{h} + \mathbf{A}^\top \bm{c} - \bm{K}^\top \bm{w} = \mathbf{0}, \\
    \quad & \bm{c} \geq \mathbf{0}. \nonumber
\end{align}
\noindent Similarly, to find $\bar{\theta}$, the dual optimization problem is formulated as
\begin{align} \label{opt:dual_upper}
    \underset{\bm{w}, \bm{c}}{\text{minimize}} \quad & \bm{w}^\top \mathbf{y} + 
    \psi_{1 - \alpha / 2}
    \lVert \bm{w} \rVert_2 + \mathbf{b}^\top \bm{c} \nonumber \\
    \textrm{subject to} \quad & \bm{h} - \mathbf{A}^\top \bm{c} - \bm{K}^\top \bm{w} = \mathbf{0}, \\
    \quad & \bm{c} \geq \mathbf{0}. \nonumber
\end{align}
We will see that these dual optimization problems provide useful insights on the intervals that subsequently lead to the construction of the Prior-Optimized intervals as discussed next.

\subsection{Prior-Optimized Intervals (PO)} \label{sec:po_intervals}
As noted in \cite{patil}, considering the dual optimization problems \eqref{opt:dual_lower} and \eqref{opt:dual_upper} can provide useful insight into the interval $[ \barbelow{\theta}_{\mathrm{OSB}}, \bar{\theta}_{\mathrm{OSB}} ]$. Namely, let $( \barbelow{\bm{w}}, \barbelow{\bm{c}})$ and $( \bar{\bm{w}}, \bar{\bm{c}})$ be dual variables satisfying the constraints in the optimization problems \eqref{opt:dual_lower} and \eqref{opt:dual_upper}, respectively. Further, consider an interval of the form
\begin{equation} \label{int:it_would_be_great}
    \big[ \barbelow{\bm{w}}^\top \mathbf{y} - z_{1 - \alpha / 2} \lVert \barbelow{\bm{w}} \rVert_2  - \mathbf{b}^\top \barbelow{\bm{c}}, \bar{\bm{w}}^\top \mathbf{y} + z_{1 - \alpha / 2} \lVert \bar{\bm{w}} \rVert_2 + \mathbf{b}^\top \bar{\bm{c}} \big].
\end{equation}
Patil et al.\ demonstrate in \cite{patil} that this interval has the correct coverage, i.e., it satisfies Eq.~\eqref{eq:correctCoverage}, for any fixed $(\barbelow{\bm{w}}, \barbelow{\bm{c}})$ and $(\bar{\bm{w}}, \bar{\bm{c}})$ that satisfy the dual constraints. However, as \cite{patil} also notes (an echo of Tenorio et al.\ \cite{tenorio}), the optimized dual variables depend on $\mathbf{y}$ and therefore an optimized version of interval Eq.~\eqref{int:it_would_be_great} similar to problems \eqref{opt:dual_lower} and \eqref{opt:dual_upper} does not necessarily satisfy the coverage requirement (which is why problems \eqref{opt:dual_lower} and \eqref{opt:dual_upper} replace $z_{1 - \alpha / 2}$ by $\psi_{1 - \alpha / 2}$ to inflate the interval).

Although this subtle point appears to block a path towards provable coverage for the optimized interval, the coverage guarantee with respect to fixed elements satisfying the dual constraints exposes an opportunity. Namely, since we may choose any $( \barbelow{\bm{w}}, \barbelow{\bm{c}})$ and $( \bar{\bm{w}}, \bar{\bm{c}})$ satisfying the dual constraints and not depending on~$\mathbf{y}$, then perhaps we can use extra information to optimally choose these variables. This perspective drives the ethos of the PO intervals.

To choose optimal $( \barbelow{\bm{w}}, \barbelow{\bm{c}})$ and $( \bar{\bm{w}}, \bar{\bm{c}})$ (that do not depend on $\mathbf{y}$), we use a decision-theoretic framework. We follow the language and notation from Berger~\cite{berger_dt}. First, define the action space to consist of real number pairs, $\mathcal{A} = \{ (a, b) \; | \; a, b \in \mathbb{R}\}$. Then, a decision rule, $\delta : \mathbb{R}^m \rightarrow \mathcal{A}$, is a function mapping an observation to an element in the action space, i.e., $\delta(\mathbf{y}) = (\barbelow{\theta}(\mathbf{y}), \bar{\theta}(\mathbf{y}))$. Let $\mathcal{D}$ be the space of all such decision rules. To match the form of the interval in Eq.~\eqref{int:it_would_be_great}, we parameterize the decision rule
\smash{$\delta(\mathbf{y}; \barbelow{\bm{w}}, \barbelow{\bm{c}}, \bar{\bm{w}}, \bar{\bm{c}}) = \big( \barbelow{\theta}(\mathbf{y}; \barbelow{\bm{w}}, \barbelow{\bm{c}}), \bar{\theta}(\mathbf{y}; \bar{\bm{w}}, \bar{\bm{c}}) \big)$}, where
\begin{align}
    \barbelow{\theta}(\mathbf{y}; \barbelow{\bm{w}}, \barbelow{\bm{c}}) &= \barbelow{\bm{w}}^\top \mathbf{y} - z_{1 - \alpha / 2} \lVert \barbelow{\bm{w}} \rVert_2  - \mathbf{b}^\top \barbelow{\bm{c}}, \\
    \bar{\theta}(\mathbf{y}; \bar{\bm{w}}, \bar{\bm{c}}) &= \bar{\bm{w}}^\top \mathbf{y} + z_{1 - \alpha / 2} \lVert \bar{\bm{w}} \rVert_2 + \mathbf{b}^\top \bar{\bm{c}}.
\end{align}
Note, these endpoints are affine functions of the data, hence we are limiting ourselves to affine decision rules. For compactness, we simply write
$\underline{\theta}(\mathbf{y})$
for
$\underline{\theta}(\mathbf{y}; \underline{\bm{w}}, \underline{\bm{c}})$
and
$\bar{\theta}(\mathbf{y})$ for $\bar{\theta}(\mathbf{y}; \bar{\bm{w}}, \bar{\bm{c}})$. To guarantee the coverage of the intervals chosen by decision rules parameterized in this way, we define the decision space on which coverage is guaranteed by
\begin{equation} \label{eq:coverage_decision_space}
    \mathcal{D}_c := \left\{ \delta \in \mathcal{D} \; | \; \bm{h} + \mathbf{A}^\top \barbelow{\bm{c}} - \bm{K}^\top \barbelow{\bm{w}} = \mathbf{0}, \bm{h} - \mathbf{A}^\top \bar{\bm{c}} - \bm{K}^\top \bar{\bm{w}} = \mathbf{0}, \, \barbelow{\bm{c}}, \bar{\bm{c}} \geq \mathbf{0} \right\}.
\end{equation}
Note, not all elements of the action space $\mathcal{A}$ define intervals with $a < b$, and by extension, the decision rules in $\mathcal{D}_c$ are not guaranteed to produce intervals for all realizations of data. However, all the decision rules are guaranteed (by their construction) to cover the true functional value $\theta$ at least $(1 - \alpha)$ percent of the time. Since a decision rule that covers the truth must be an interval, this means that for every $\delta \in \mathcal{D}_c$, $\delta$ must produce an interval at least $(1 - \alpha)$ percent of the time. Empirically, in all simulations for the applications in Section~\ref{sec:story_section} and Section~\ref{sec:particle_spectrum}, we did not encounter a situation in which $\barbelow{\theta}(\mathbf{y}) > \bar{\theta}(\mathbf{y})$. However, in principle, since $\mathbf{y} \sim \mathcal{N}(\bm{K} \blambda, \mathbf{I})$, which has positive measure on all subsets of $\mathbb{R}^m$, there exists a realization $\mathbf{y}'$ of the data such that $\barbelow{\theta}(\mathbf{y}') > \bar{\theta}(\mathbf{y}')$. We refer to this as the pathological case.

Picking an optimal decision first requires a way to measure the quality of the decision rule via a loss function $L: \mathcal{A} \rightarrow \mathbb{R}$. Since we only consider $\delta \in \mathcal{D}_c$, i.e., decision rules for which coverage is guaranteed, we restrict our loss function definition to only account for interval size. As such, we simply define the loss to be the interval width, i.e.,
\begin{equation} \label{def:loss_function}
    L(\delta(\mathbf{y})) =  \bar{\theta}(\mathbf{y}) - \underline{\theta}(\mathbf{y}) = \left(\bar{\bm{w}} - \underline{\bm{w}} \right)^\top \mathbf{y} + z_{1 - \alpha / 2} \left( \lVert \bar{\bm{w}} \rVert_2 + \lVert \underline{\bm{w}} \rVert_2 \right) + \mathbf{b}^\top \left(\bar{\bm{c}} + \underline{\bm{c}} \right).
\end{equation}
Note, this definition of interval size is also the Lebesgue measure of the interval.

The \emph{risk functional} of a decision rule is then defined as the expectation of the loss function with respect to the probability measure on the data.
In other words,
\begin{equation}
    R(\delta) = \mathbb{E}_{\mathbf{y}} \left[ L(\delta(\mathbf{y})) \right] = \left(\bar{\bm{w}} - \underline{\bm{w}} \right)^\top \mathbf{K} \blambda + z_{1 - \alpha / 2} \left( \lVert \bar{\bm{w}} \rVert_2 + \lVert \underline{\bm{w}} \rVert_2 \right) + \mathbf{b}^\top \left(\bar{\bm{c}} + \underline{\bm{c}} \right).
\end{equation}
Optimal choices of the decision rule can now be defined in terms of this risk functional.

Note that $R(\delta)$ is a function of the true parameter $\blambda$. Luckily, in physical science applications, we often have access to well-justified prior information about $\blambda$ stemming from theory or previous experimentation. As such, the Bayes Risk Principle sensibly defines an optimal decision rule (see Section 1.5 of Berger \cite{berger_dt}). Given a prior distribution $\pi_{\blambda}$ on $\blambda$ with expectation $\bm{m}_{\bm{\lambda}}$, the \emph{Bayes risk} (Definition 6 in Section 1.3.2 of~\cite{berger_dt}) of a decision rule $\delta$ is~defined,
\begin{equation} \label{def:bayes_risk}
    r(\pi_{\blambda}, \delta) := \mathbb{E}_{\blambda} \left[ R(\delta) \right] = \left(\bar{\bm{w}} - \underline{\bm{w}} \right)^\top \mathbf{K} \bm{m}_{\blambda} + z_{1 - \alpha / 2} \left( \lVert \bar{\bm{w}} \rVert_2 + \lVert \underline{\bm{w}} \rVert_2 \right) + \mathbf{b}^\top \left(\bar{\bm{c}} + \underline{\bm{c}} \right).
\end{equation}
Under the Bayes Risk Principle, an optimal decision rule minimizes the Bayes risk and is called a \emph{Bayes rule}. As such, we define the Prior-Optimized intervals by the decision rule that is the Bayes rule under the above construction, i.e., we find $\delta^* \in \mathcal{D}_c$ such that
\begin{equation}
    r(\pi_{\blambda}, \delta^*) \leq r(\pi_{\blambda}, \delta), \quad \text{for all } \delta \in \mathcal{D}_c.
\end{equation}
Since our class of decision rules is parameterized by the dual variables, $(\barbelow{\bm{w}}, \barbelow{\bm{c}}, \bar{\bm{w}}, \bar{\bm{c}})$, finding the Bayes decision rule is achieved by solving the following optimization problem:
\begin{align} \label{int:po}
    \underset{\barbelow{\bm{w}}, \barbelow{\bm{c}}, \bar{\bm{w}}, \bar{\bm{c}}}{\text{minimize}} \quad & r(\pi_{\blambda}, \delta) \nonumber \\
    \textrm{subject to} \quad & \bm{h} + \mathbf{A}^\top \barbelow{\bm{c}} - \bm{K}^\top \barbelow{\bm{w}} = \mathbf{0}, \\
    \quad & \bm{h} - \mathbf{A}^\top \bar{\bm{c}} - \bm{K}^\top \bar{\bm{w}} = \mathbf{0}, \nonumber \\
    \quad & \barbelow{\bm{c}}, \bar{\bm{c}} \geq \mathbf{0}. \nonumber
\end{align}
Under the above construction, only the prior expectation, $\bm{m}_{\blambda}$ needs to be specified, and the coverage guarantee follows from the constraints inherited from optimization problems~\eqref{opt:dual_lower} and \eqref{opt:dual_upper}. The optimization problem \eqref{int:po} is a convex optimization problem,
and can be cast as a second-order cone program and solved efficiently.
Note, to solve optimization~\eqref{int:po} in practice, we can equivalently maximize the expected lower endpoint and minimize the expected upper endpoint.

It is important to emphasize that since interval~\eqref{int:it_would_be_great} has provable coverage for fixed dual variables and since here the choice of these variables is done without using the data~$\mathbf{y}$, it follows that the PO interval also has provable coverage. This holds true even in the presence of affine constraints on $\blambda$ and for column-rank-deficient $\mathbf{K}$. Importantly, despite the use of the prior $\pi_{\blambda}$, the coverage guarantee is entirely frequentist. The prior is only used to optimize the interval width. If the prior is misspecified (as it usually is), the width might be suboptimal, but the coverage guarantee still holds. This is different from standard Bayesian use of prior information in which case the coverage depends strongly on the choice of the prior \cite{patil}.

Additionally, although we have made specific choices regarding the parameterization of the decision rule and the loss function, the above decision-theoretic framework is general enough to accommodate a variety of choices and modifications. For instance, one might parameterize the decision rule with non-linear endpoints. Or, with access to a prior covariance matrix, one might choose a loss function to incorporate this second-order information in addition to the information on the first moment. More generally, this framework suggests a meta-algorithm to find interval estimators with frequentist coverage guarantees. Namely, if one is able to define a set of decision rules for which coverage is guaranteed as we did in Eq.~\eqref{eq:coverage_decision_space}, then the Bayes Risk Principle can be used to optimize the expected interval size.

\subsection{Other Related Intervals} \label{sec:other_intervals}
The model described by Eq.~\eqref{eq:normal_stat_model} is a well-studied statistical model. In particular, the functional interval estimation task accomplished by the OSB and PO intervals can also be accomplished by intervals based on the least-squares estimator, minimax intervals \cite{donoho_94}, and simultaneous strict bounds intervals (henceforth referred to as ``SSB'' intervals) \cite{stark_strict_bounds}. Each alternative method provides a reference point against which OSB and PO intervals can be compared.

Each of these methods can be categorized according to its assumptions and properties. For instance, both the least-squares and minimax intervals require a full-column-rank linear model to ensure that the model has a trivial nullspace. Otherwise, the interval endpoints can be unbounded depending on the orientation of the model's null space. Additionally, these intervals can be categorized according to whether they are designed to provide one-at-a-time functional coverage or simultaneous coverage. The SSB intervals are the only simultaneous intervals we consider here. If considered for a single functional, this criterion makes these intervals more conservative than the others, since their coverage is required to hold over a set of functionals, as opposed to one. As such, we expect these intervals to be wider than both the OSB and PO intervals. Similarly, the minimax intervals are, by their nature, conservative, and thus, we also expect these intervals to be wider than both the OSB and PO intervals.

Perhaps the most obvious interval construction to use with model in Eq.~\eqref{eq:normal_stat_model} is the unconstrained least-squares (LS) intervals. Not only are these intervals the obvious procedure to use given the statistical model of the problem, but \cite{patil} shows that the OSB intervals for unconstrained full-column-rank models are equivalent to least-squares intervals, and thus are a natural comparison. We refer readers to Appendix C of \cite{patil} for a full description and derivation of this result.

The primary necessary condition for constructing finite intervals using the least-squares estimator is for $\bm{K}$ to have full column rank, which implies the invertibility of $\bm{K}^T \bm{K}$. Note, as shown in Eq.~\eqref{eq:cholesky_transform_data}, the data model can always be transformed to have identity covariance matrix. As such, we assume identity covariance. Assuming this condition holds, we can compute the minimum variance unbiased estimator of the parameter, $\hat{\blambda} = \left( \bm{K}^\top \bm{K} \right)^{-1} \bm{K}^\top \mathbf{y}$, and construct a confidence interval with guaranteed coverage $\left(1 - \alpha \right)$ for the quantity of interest $\theta = \bm{h}^\top \blambda$ as the interval:
\begin{equation} \label{eq:least_squares_interval}
    \big[\barbelow{\theta}_{\mathrm{LS}}, \bar{\theta}_{\mathrm{LS}} \big] = \big[\bm{h}^\top \hat{\blambda} - \gamma, \bm{h}^\top \hat{\blambda} + \gamma \big]
\end{equation}
where $\gamma = z_{1 - \alpha / 2} (\bm{h}^T \left( \bm{K}^\top \bm{K} \right)^{-1}\bm{h})^{1/2}$.

By contrast, as shown in Sections \ref{sec:story_section} and \ref{sec:particle_spectrum}, the SSB intervals \cite{stark_strict_bounds} are finite even in rank-deficient situations, but are overly conservative for a single functional given their simultaneous coverage guarantee. These intervals are computed similarly to optimization problems \eqref{opt:osb_lower} and \eqref{opt:osb_upper}, where
\begin{equation}
    \big[ \barbelow{\theta}_{\mathrm{SSB}}, \bar{\theta}_{\mathrm{SSB}} \big] = \bigg[ \underset{\blambda \in \mathcal{Z}(\chi^2_{n, 1 - \alpha})}{\text{min}} \bm{h}^\top \blambda, \underset{\blambda \in \mathcal{Z}(\chi^2_{n, 1 - \alpha})}{\text{max}} \bm{h}^\top \blambda \bigg].
\end{equation}
Note, the SSB intervals are constructed by replacing the feasible region $\mathcal{Z}(\psi_{1 - \alpha/2}^2 )$ from the OSB interval construction in Eq.~\eqref{eq:osb_construct} with $\mathcal{Z}(\chi^2_{n, 1 - \alpha})$, where $\chi^2_{n, 1 - \alpha}$ is the $(1 - \alpha)$-th quantile of the chi-squared distribution with $n$ degrees of freedom  \cite{patil, stark_strict_bounds}.

The half-width of the fixed-width affine minimax confidence intervals is described in \cite{donoho_94}, using the notation $C^*_{\alpha, A}(\sigma)$, where $\alpha$ is the same as above, and $\sigma$ is the standard deviation of the Gaussian noise. Computing this exact half-width is difficult, so we instead note that Corollary 2 in \cite{donoho_94} provides upper and lower bounds, which can be more easily computed. Following \cite{stark_geomag}, we compute the lower and upper bounds of the minimax interval half-width by evaluating the modulus of continuity as an optimization problem. We use the double of these bounds to compare against the other interval widths. Generally, given $\eps > 0$, the modulus of continuity is defined in \cite{donoho_94} as follows:
\begin{equation} \label{def:modulus_continuity}
    \omega \left( \eps; \bm{h}, \bm{K}, \bm{\Lambda} \right) = \sup_{\bm{\lambda}_1,\bm{\lambda}_{-1}} \big\{ \lvert \bm{h}^\top \bm{\lambda}_1 - \bm{h}^\top \bm{\lambda}_{-1} \rvert: \lVert \bm{K}(\bm{\lambda}_1 - \bm{\lambda}_{-1}) \rVert_2 \leq \eps, \, \bm{\lambda}_1, \bm{\lambda}_{-1} \in \bm{\Lambda} \big\},
\end{equation}
where $\bm{h}$ is the functional of interest, $\bm{K}$ is the smearing matrix, and $\bm{\Lambda}$ is a convex subset of the underlying parameter space. For instance, in Section~\ref{sec:story_section}, we consider the case when this subset is simply the positive orthant of $\mathbb{R}^n$. Using the modulus of continuity, Corollary 2 in \cite{donoho_94} gives the following upper and lower bounds:
\begin{equation} \label{eq:donoho_lower_upper_bounds}
    \omega ( 2 z_{1 - \alpha} \sigma ) \leq C^*_{\alpha, A}(\sigma) \leq \omega (2 z_{1 - \alpha / 2} \sigma )
\end{equation}
where $z_{1 - \alpha}$ and $z_{1 - \alpha / 2}$ are the $(1 - \alpha)$ and $(1 - \alpha / 2)$ level quantiles of a standard normal distribution. Hence, we can find the lower bound indicated in Eq.~\eqref{eq:donoho_lower_upper_bounds} using the following optimization problem:
\begin{align} \label{opt:donoho_lower}
    \underset{\bm{\lambda}_1, \bm{\lambda}_{-1}}{\text{maximize}} \quad & \lvert \bm{h}^\top \bm{\lambda}_1 - \bm{h}^\top \bm{\lambda}_{-1} \rvert \nonumber \\
    \textrm{subject to} \quad &\lVert \bm{K} \bm{\lambda}_1 - \bm{K} \bm{\lambda}_{-1} \rVert_2 \leq 2 z_{1 - \alpha} \sigma,  \\
    & \mathbf{A} \bm{\lambda}_{-1} \leq \mathbf{b}, \nonumber \\
    & \mathbf{A} \bm{\lambda}_1 \leq \mathbf{b}. \nonumber
\end{align}
To cast optimization problem \eqref{opt:donoho_lower} into a more standard form, define $\mathbf{z}~:=~\begin{bsmallmatrix} \bm{\lambda}_1^\top & \bm{\lambda}_{-1}^\top \end{bsmallmatrix}^\top~\in~\mathbb{R}^{2n}$, $\mathbf{B} := \begin{bsmallmatrix} \mathbf{I}_n & -\mathbf{I}_n \end{bsmallmatrix} \in \mathbb{R}^{n \times 2n}$, $\mathbf{f}^\top := \bm{h}^\top \mathbf{B} \in \mathbb{R}^{1 \times 2n}$,
$\mathbf{A}_2~:=~\begin{bsmallmatrix} \mathbf{A} & \mathbf{0}_{q \times 2n} \\ \mathbf{0}_{q \times 2n} & \mathbf{A}\end{bsmallmatrix} \in \mathbb{R}^{2q \times 2n}$,
$\mathbf{b}_2~:=~\begin{bsmallmatrix} \mathbf{b} \\ \mathbf{b} \end{bsmallmatrix} \in \mathbb{R}^{2q \times 1}$,
and $\Tilde{\bm{K}} := \bm{K} \mathbf{B}$, where $\mathbf{I}_n$ 
is the $n \times n$ identity matrix. Thus, we can re-write Eq.~\eqref{opt:donoho_lower} as follows:
\begin{align} \label{opt:donoho_lower_standard}
    \underset{\mathbf{z} \in \mathbb{R}^{2n}}{\text{maximize}} \quad & \lvert \mathbf{f}^\top \mathbf{z} \rvert \nonumber \\
    \textrm{subject to} \quad &\lVert \Tilde{\bm{K}} \mathbf{z} \rVert_2 \leq 2 z_{1 - \alpha} \sigma,  \\
    & \mathbf{A}_2 \mathbf{z} \leq \mathbf{b}_2 \nonumber.
\end{align}
Note, when solving optimization problem~\eqref{opt:donoho_lower_standard} with a non-negativity constraint on the parameters as in Section~\ref{sec:story_section}, we can drop the absolute value signs in the objective function. Since the indexing of $\bm{\lambda}_1$ and $\bm{\lambda}_{-1}$ is arbitrary, if the objective function is negative, the labels can simply be reversed to render it positive. Thus, optimization problem~\eqref{opt:donoho_lower_standard} simply becomes a quadratically constrained linear program which we can solve using standard convex optimization algorithms. Computing the minimax half-width upper bound is performed almost exactly as above in the optimization problem \eqref{opt:donoho_lower_standard}, but by replacing $z_{1 - \alpha}$ with $z_{1 - \alpha / 2}$.

As shown in the results section below, full-column-rank $\bm{K}$ produces finite lower and upper bounds on the minimax interval width. However, in the rank-deficient cases, since our functionals of interest are not in the row space of $\bm{K}$ (i.e., they are not in the orthogonal complement of the null space), the lower bounds on the minimax interval widths are infinite. As a result, these minimax intervals are not practical for arbitrary functionals in the rank-deficient situation. Our discussion of minimax intervals is restricted to the notion of minimax optimality for fixed-width intervals as defined in~\cite{donoho_94}. It is worth noting that there are other notions of minimax optimality such as that of \cite{evans_hansen_stark, schafer_stark}.

\section{Application to Wide-Bin Deconvolution} \label{sec:story_section}

\subsection{Simulation Setup}
In this section, we describe the simulation setup we use to numerically explore the coverage and width of the OSB and PO intervals presented above. We must define a true intensity function $f$, a smearing kernel $k$, and a collection of functionals for which we seek to build confidence intervals.

To define the intensity function, we utilize Theorem 1.2.1 in \cite{reiss}, which gives us a convenient way to define and sample from a Poisson point process. As described in Section~\ref{sec:unfolding_problem}, let $X$ represent a random collision event in the true space $T$. Let $\tau \sim \textrm{Poisson}(\lambda(T))$ be independent from $X \sim P_X$, where the distribution $P_X$ is the same as the normalized measure $\lambda / \lambda(T)$. Since the mean of a Poisson random vriable is the same as its rate parameter, $\lambda(T) = \mathbb{E}[\tau]$, and hence when the densities exist, we have
\begin{equation} \label{eq:intensity_construction}
    f(x) = \mathbb{E}[\tau] p_X(x)
\end{equation}
where $p_X$ is the Radon--Nikodym derivative of $P_X$ (for more details about this construction, see \cite{kuusela_masters_thesis}). Practically, Eq.~\eqref{eq:intensity_construction} allows us to construct an intensity function for the true particle-level Poisson point process by specifying an expected number of events $\mathbb{E}[\tau]$ over the state space $T$, and a probability density $p_X$.

To define the density $p_X$, we use a Gaussian mixture model (GMM). For $x \in \mathbb{R}$, we let $\mathcal{N}(x; \mu, \sigma^2)$ denote the probability density function of a Gaussian distribution with mean $\mu$ and variance $\sigma^2$ at $x$. For our simulations, we define $p_X$ as follows
\begin{equation}
    p_X(x) = \pi_1 \mathcal{N}(x; \mu_1, \sigma_1^2) + \pi_2 \mathcal{N}(x; \mu_2, \sigma_2^2),
\end{equation}
where $\pi_i = \mathbb{P}[Z = i]$, where $Z \in \{ 1, 2 \}$ and $p_X(x) = \sum_{z = 1}^2 p_{X, Z}(x, z)$, where $p_{X, Z}$ is the joint density of $(X, Z)$. Hence, the intensity $f$ is defined as
\begin{equation} \label{eq:simulation_intensity}
    f(x) = N_c \big\{ \pi_1 \mathcal{N}(x; \mu_1, \sigma_1^2) + \pi_2 \mathcal{N}(x; \mu_2, \sigma_2^2) \big\}
\end{equation}
\noindent where $N_c = \mathbb{E}[\tau]$. The full list of parameter settings used for the GMM simulations is shown in Table~\ref{table:parameter_settings}.
\begin{table}[t]
    \centering
    \caption{Parameter settings for the GMM simulation.}
    \begin{tabular}{ccc}
         \hline
         Parameter & Value & Description \\
         \hline 
         $N_c$ & $10{,}000$ & Expected number of collision events over the entire true space \\
         $(\mu_1, \mu_2)$ & $(-2, 2)$ & Mixture means \\
        $(\sigma_1^2, \sigma_2^2)$ & $(1, 1)$ & Mixture variances \\
        $T$ ($S$) & $[-7, -7]$ & True and smeared spaces \\
        $(\pi_1, \pi_2)$ & $(0.3, 0.7)$ & Mixture weights \\
        \hline
    \end{tabular}
    \label{table:parameter_settings}
\end{table}
These parameter settings are chosen to continue and build upon the simulation setup described in Section 3.4.1 of \cite{kuusela_phd_thesis}. Note, in our setup, we use only two mixture components without the uniform background.

We define the smearing kernel to correspond to the situation where we add independent Gaussian noise to smear the points drawn from the intensity in Eq.~\eqref{eq:simulation_intensity}. Namely,
\begin{equation}
    k(s, t) = \frac{1}{\sqrt{2\pi}\gamma} \exp \left(-\frac{1}{2\gamma^2} (s - t)^2 \right).
\end{equation}
Said differently, a true point $X$ creates an observed point $Y = X + \epsilon$, where $\epsilon \sim \mathcal{N}(0, \gamma^2)$ and $X$ and $\epsilon$ are independent. As noted above, since $T = S = [-7, 7]$, points that are smeared beyond these boundaries are not included among the observed points. This kernel includes the parameter $\gamma$, which essentially controls the magnitude of the smearing, and hence, controls the extent to which this problem is ill-posed. For our simulations, we use $\gamma = 0.35$.

\begin{figure}
    \centering
    \begin{subfigure}[b]{0.49\textwidth}
    	\centering
    	\includegraphics[width=\textwidth]{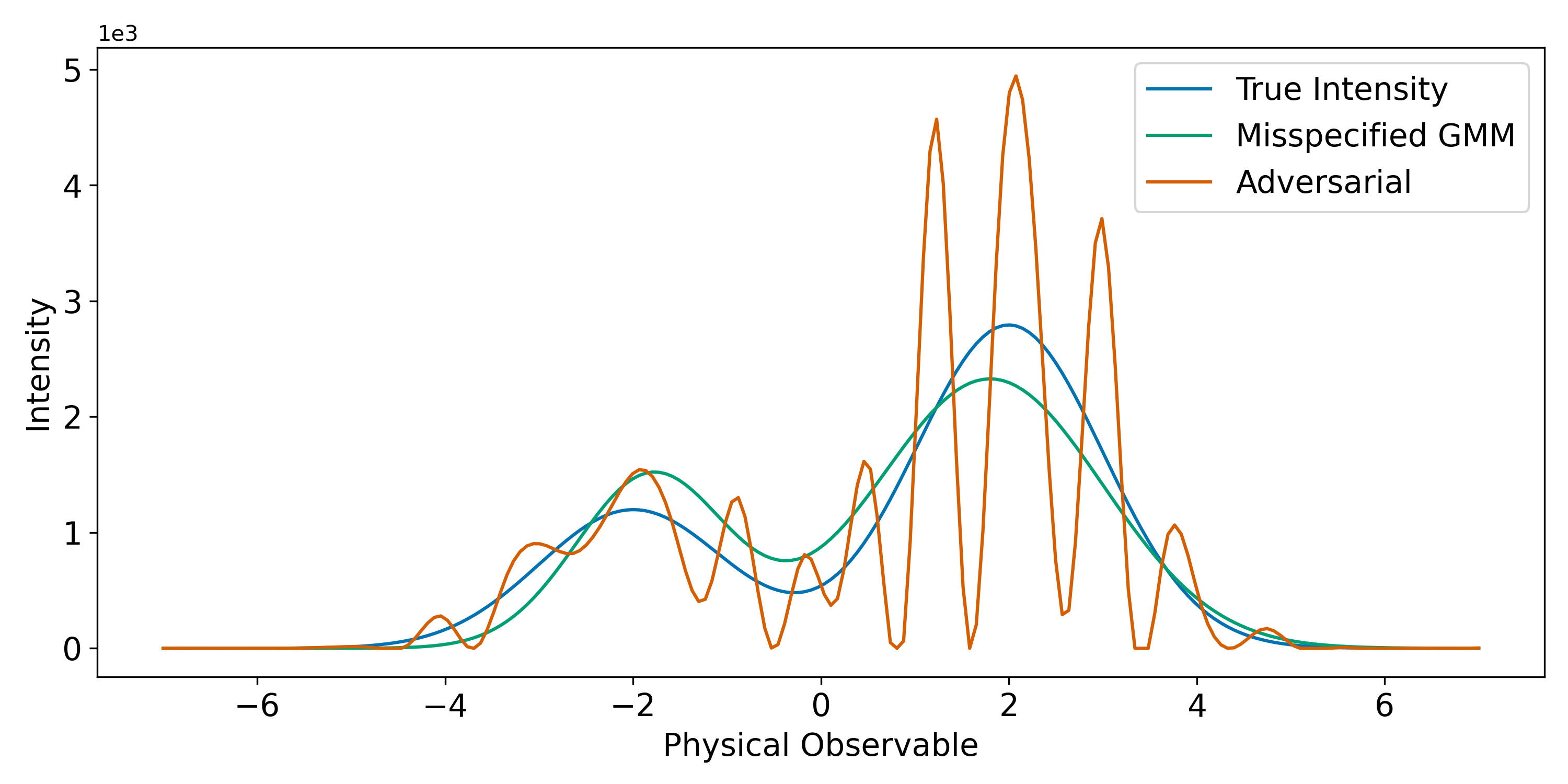}
	\caption{}
    	\label{fig:intensity_functions}
    \end{subfigure}
    \hfill
    \begin{subfigure}[b]{0.49\textwidth}
    	\centering
	\includegraphics[width=\textwidth]{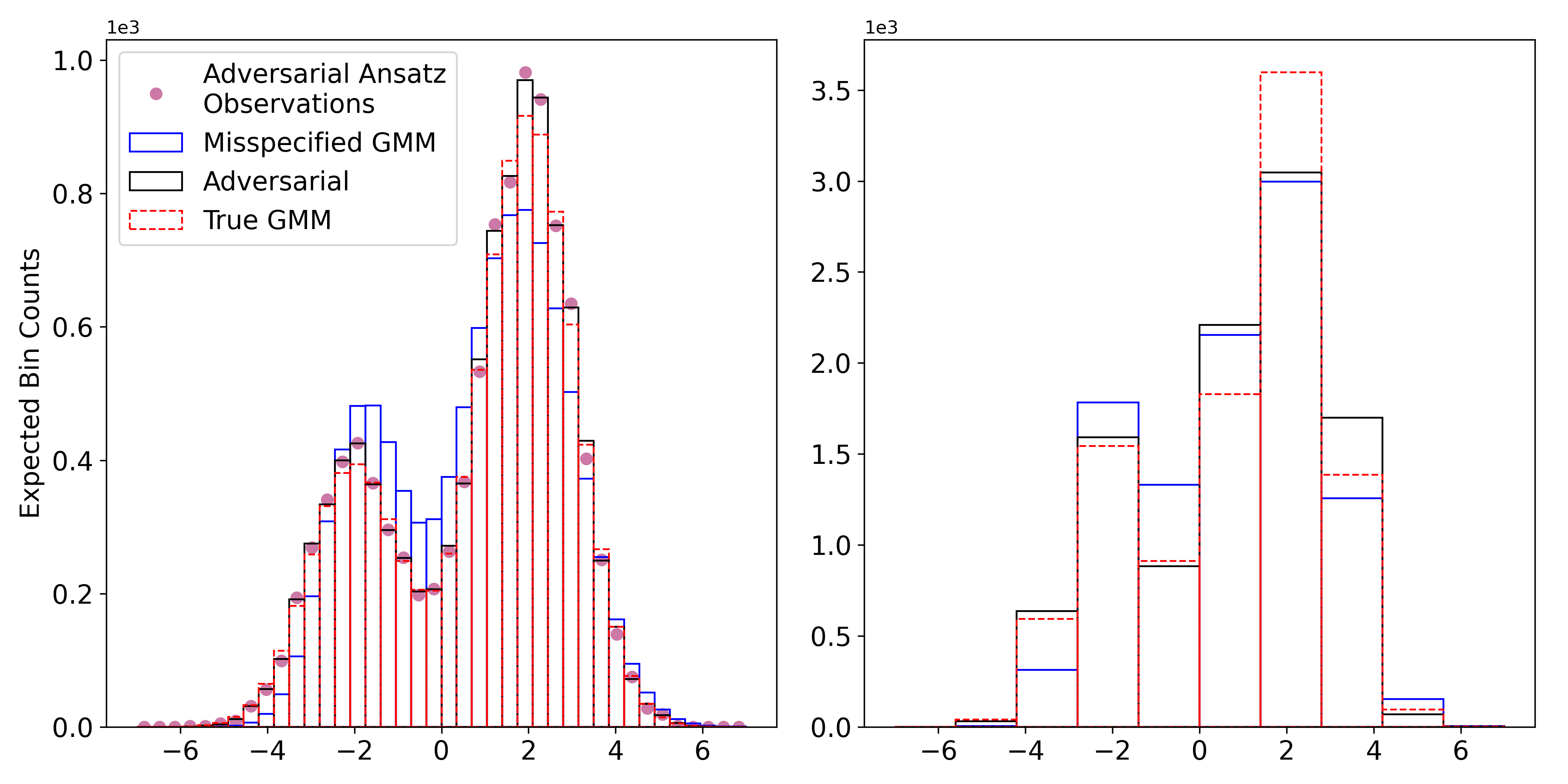}
	\caption{}
    	\label{fig:wide_bin_expectations}
    \end{subfigure}
    \caption{Fig.~\ref{fig:intensity_functions}: Illustration of the true intensity function used for simulations and the two ansatz functions used for computing the smearing matrix~$\bm{K}$. Fig.~\ref{fig:wide_bin_expectations}: \textbf{(Left)} True bin expected counts. The deviations in each bin between the true expectation and the ansatz expectations drives the systematic error.    \textbf{(Right)} Smeared bin expected counts. Note the reduced magnitude of difference in expectation with respect to the true space, emphasizing the convolutional nature of the smearing matrix.}
\end{figure}

Additionally, we create two MC ansatz functions for computing the smearing matrix~$\bm{K}$. The first is simply a slightly misspecified version of the true data generating process in Eq.~\eqref{eq:simulation_intensity}. Namely, we let $(\mu_1, \mu_2) = (-1.8, 1.8)$ and $(\sigma_1, \sigma_2) = (0.8, 1.2)$ and keep all other parts of Eq.~\eqref{eq:simulation_intensity} the same. We refer to this ansatz as the ``Misspecified GMM Ansatz''. The second is specifically created so that it fits the smeared data well but has oscillations which magnify the effect of the systematic error. To find such an ansatz, we generate a realization of data from the smeared process (i.e., use Eq.~\eqref{eq:true_poisson_data_gen} to generate a vector $\mathbf{y} \in \mathbb{R}^m$), find the non-negatively constrained least-squares estimate of the true bin counts (i.e., \smash{$\hat{\blambda} = \text{argmin}_{\blambda \geq \mathbf{0}} \lVert \mathbf{y} - \bm{K} \blambda \rVert_2$)}, and define the ansatz $f^\text{MC}$ by fitting a cubic spline to $\hat{\blambda}$ and forcing it to be non-negative. Clearly, this ansatz fits the data that generated it well. However, because it is a mildly constrained cubic spline, it is highly oscillatory. See Supplement Section~\ref{sec:appendix_a} for a more detailed account of this its creation. We refer to this ansatz as the ``Adversarial Ansatz''. Fig.~\ref{fig:intensity_functions} shows the true intensity functions and the two ansatz functions described above.

\subsection{The Wide-Bin Problem} \label{sec:wide_bin_problem}
When obtaining confidence intervals for bin counts in the unfolding problem, it is common practice to use more smeared bins than unfolded bins. Fundamentally, the ill-posedness of the problem motivates this design choice. However, as more thoroughly described in the sections below, systematic error in the Monte Carlo ansatz used when computing the smearing matrix (per Eq.~\eqref{eq:K_matrix_component}) can cause coverage issues even in simple examples. To demonstrate this point, we consider a true intensity function as described by Eq.~\eqref{eq:simulation_intensity} and construct 95\% confidence intervals for each bin in the unfolded space.

Of course, if the ansatz were correctly specified, systematic error from the wide-bin bias would not be a concern, but this would be unrealistic since it would mean assuming that we already know the unknown intensity function before carrying out the experiment. So in practice, we need to assume that the ansatz will not be correctly specified in any analysis. However, the ansatz is still likely to be fairly close to the true function since it usually follows from some assumed physical theory. As such, we first consider the Misspecified GMM Ansatz since it is nearly the same as the true intensity function.

We discretize the true space shown into $n=10$ bins, while the smeared space discretization is set to $m=40$ bins. Fig.~\ref{fig:wide_bin_expectations} shows the deviations in the expected bin counts both in the true space (left) and the smeared space (right). The differences in the left side of Fig.~\ref{fig:wide_bin_expectations} flow to the smearing matrix and provide the source for the systematic error.

\begin{figure}[htp]
    \centering
    \includegraphics[width=0.7\textwidth]{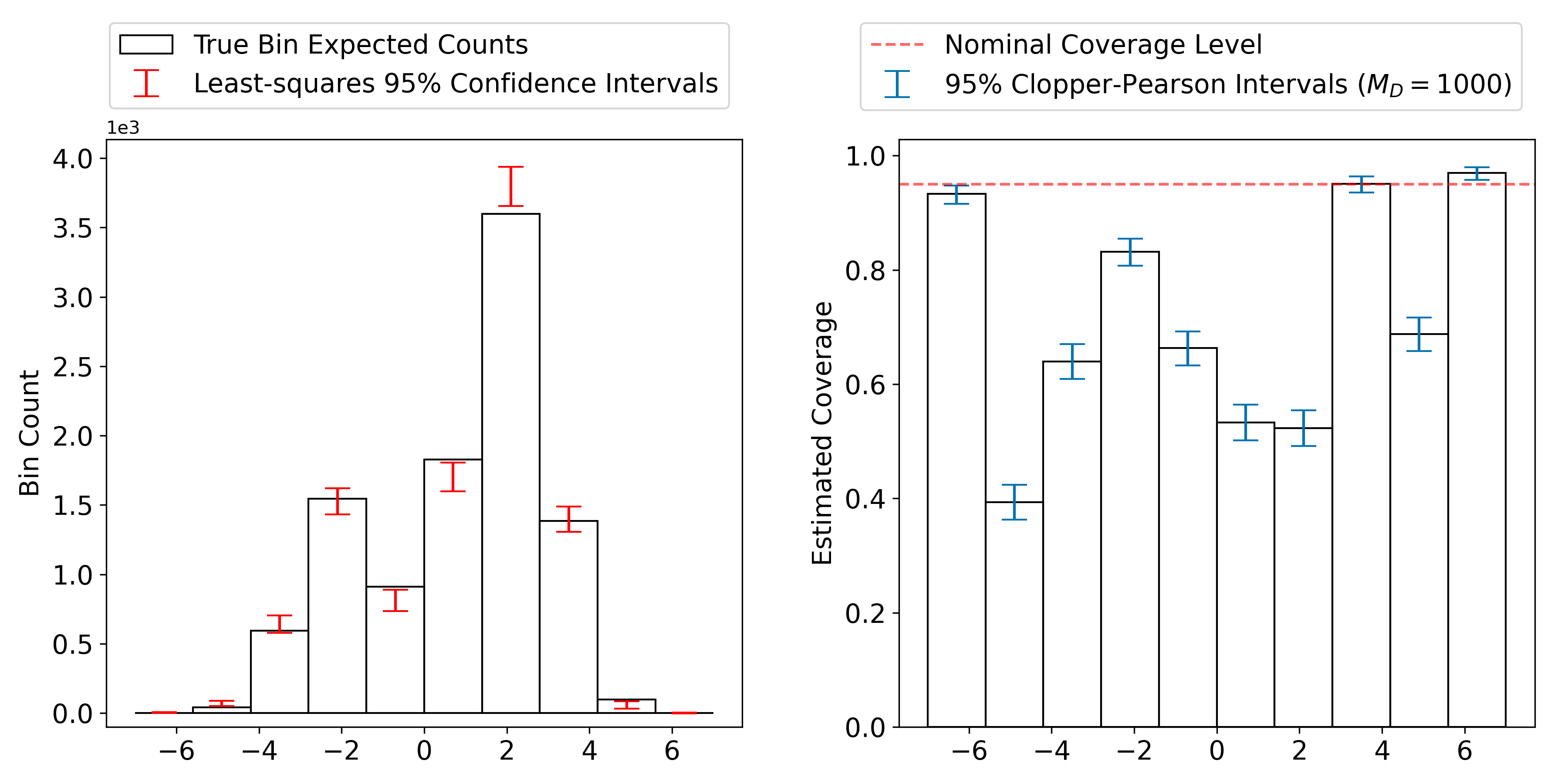}
    \caption{\textbf{(Left)} Bin count intervals constructed from one out of the $M_D$ samples of data with the Misspecified GMM Ansatz. \textbf{(Right)} Estimated bin-wise coverage of the least-squares intervals shows that most of the bin count intervals undercover.}
    \label{fig:wide_bin_ls_failure}
\end{figure}

To show how the wide-bin bias disrupts coverage guarantees with even the Misspecified GMM Ansatz, we evaluate the coverage of 95\% least-squares confidence intervals for each of the 10 unfolded bins. To estimate the coverage, we generate $M_D = 1{,}000$ realizations of smeared data, $\mathbf{y}_1, \dots, \mathbf{y}_{M_D} \sim \textrm{Poisson}(\bmu)$, where $\bmu \in \mathbb{R}^{40}$ is defined above in Eq.~\eqref{eq:means_linear_system}. Since $\bmu = \bm{K} \blambda$, where $\blambda \in \mathbb{R}^{10}$ is the vector of true bin expected counts, and the bin counts are all sufficiently large, we use the Normal approximation to the Poisson distribution to construct confidence intervals for each bin using the least-squared intervals described in Eq.~\eqref{eq:least_squares_interval}. The intervals constructed for one realization of data are shown in the left portion of Fig.~\ref{fig:wide_bin_ls_failure}. For each $i = 1, \dots, M_D$, we thus construct ten intervals. For each bin $j = 1, \dots, 10$, we estimate the coverage with the following statistic:
\begin{equation} \label{eq:estimate_binwise_coverage}
    \gamma_j = \frac{1}{M} \sum_{i = 1}^M \mathbbm{1} \left\{ \theta_{ij} \in [\barbelow{\theta}_{ij}, \bar{\theta}_{ij}] \right\}
\end{equation}
where $[\barbelow{\theta}_{ij}, \bar{\theta}_{ij}]$ is the least-squares interval computed for the $i$th realization of data for the $j$th bin,
and $\mathbbm{1}\{ A \}$ denote indicator function for event $A$. By the law of large numbers, if the procedure is working as expected, $\gamma_j \overset{P}{\to} 1 - \alpha$ for all $j = 1, \dots, 10$. However, as we can see in the right portion of Fig.~\ref{fig:wide_bin_ls_failure}, all but three of the ten intervals exhibit severely deficient coverage. So, although the intervals have a desirable width, they do not cover the true values nearly as often as we would like because of the wide-bin bias from the misspecified MC ansatz.

\subsection{Addressing the Wide-Bin Problem} \label{sec:addressing_wide_bin_problem}

If $\bm{K}$ were correctly specified when constructing the intervals, the previous coverage problem would not exist. However, with wide true bins, the $\bm{K}$ matrix will realistically never be correctly specified. As such, bin count interval coverage breaks down even with the relatively harmless Misspecified GMM Ansatz.

\begin{figure}
	\begin{subfigure}[b]{\textwidth}
		\centering
		\includegraphics[width=0.7\textwidth]{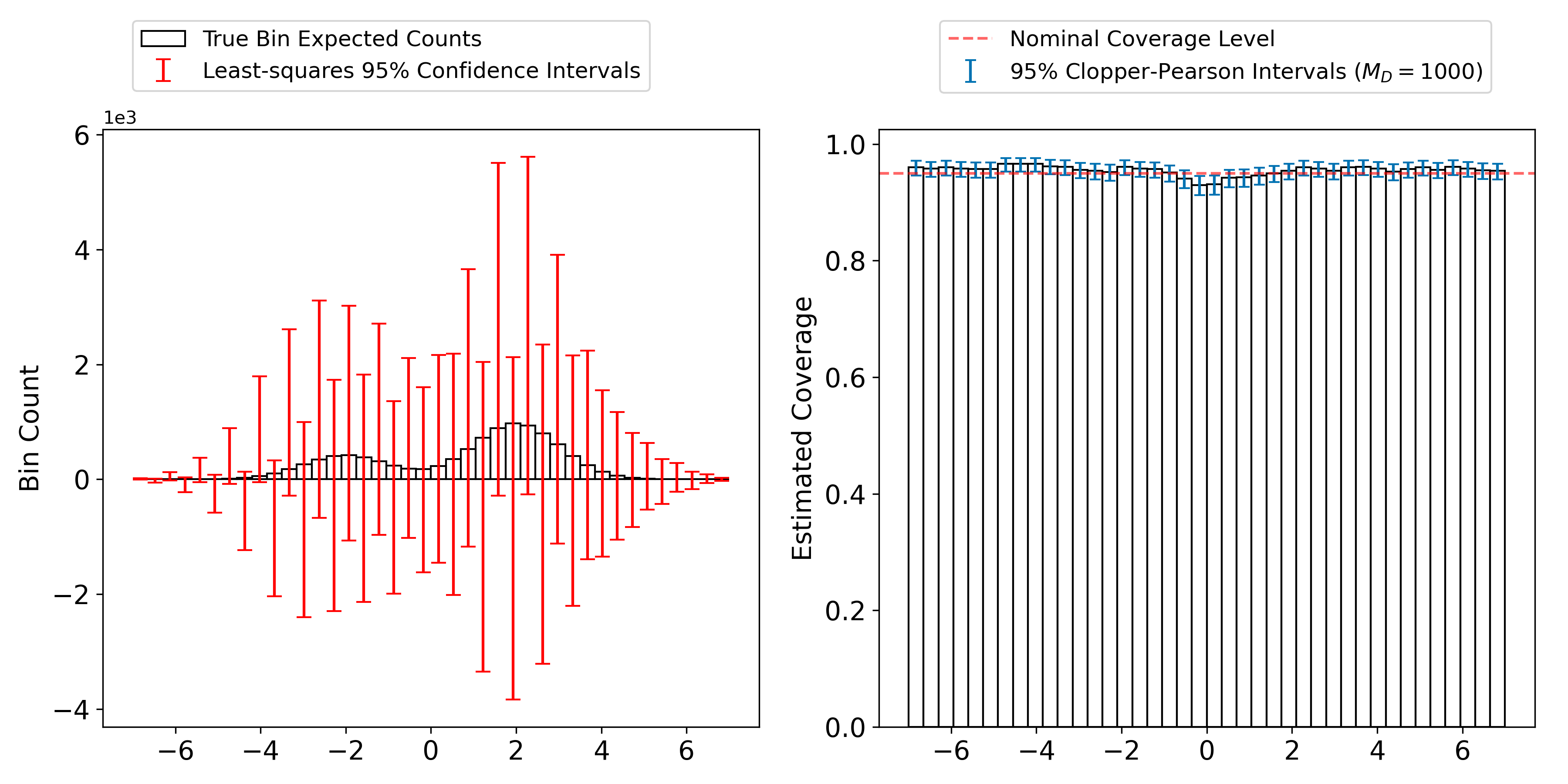}
		\caption{}
		\label{fig:fine_bin_fix}
	\end{subfigure}
	\\
	\begin{subfigure}[b]{\textwidth}
		\centering
		\includegraphics[width=0.7\textwidth]{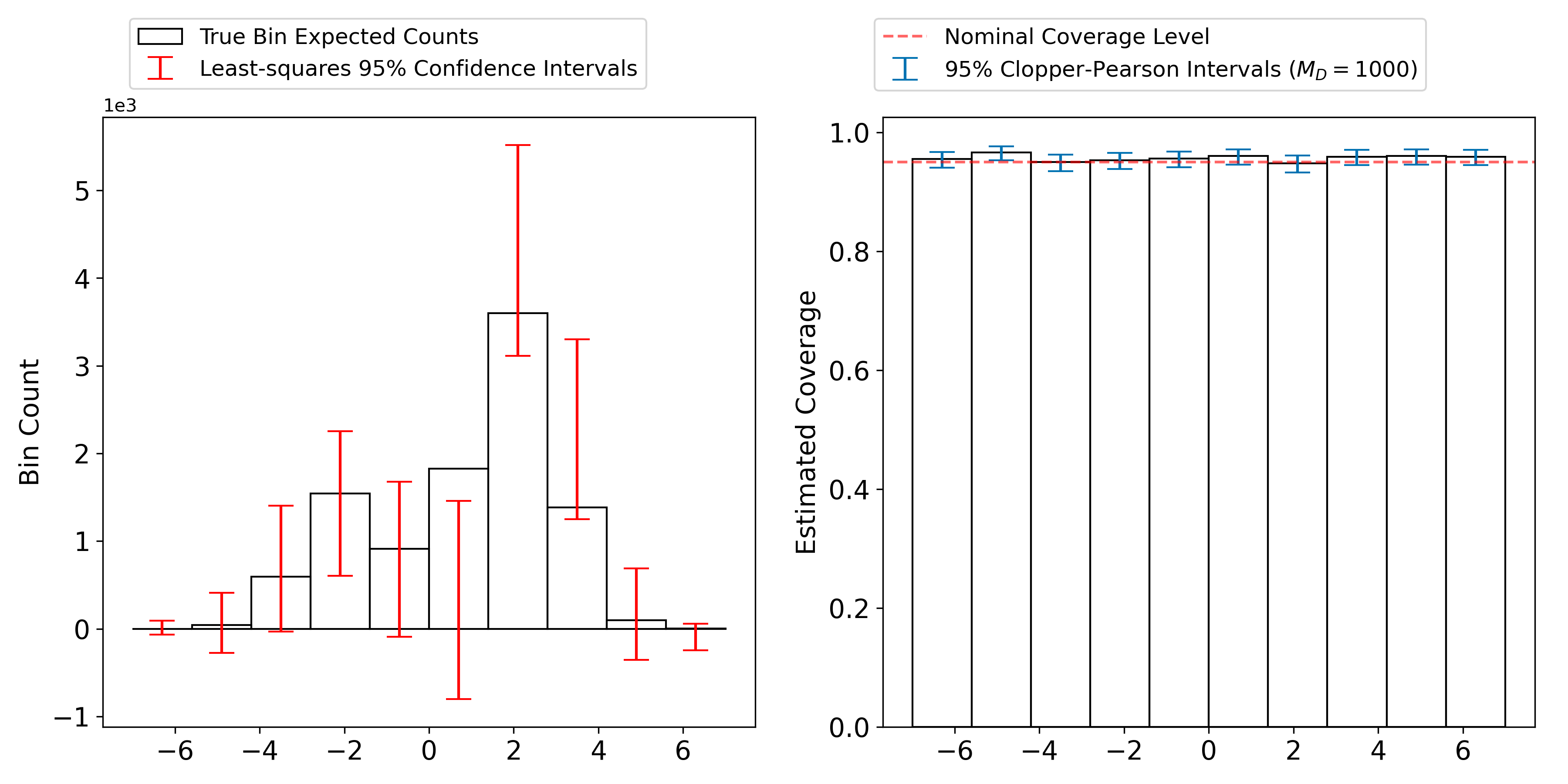}
		\caption{}
		\label{fig:agg_bin_intervals_and_coverage}
	\end{subfigure}
	\caption{\ref{fig:fine_bin_fix}: Unfolding with $n=40$ true bins and least-squares intervals. The \textbf{Left} and \textbf{Right} are analogous to those in Fig.~\ref{fig:wide_bin_ls_failure}, but with the $n=40$ true bins. Using more true bins fixes the coverage problem of the least-squares intervals in Fig.~\ref{fig:wide_bin_ls_failure} (\textbf{Right}), at the expense of significantly wider bin-wise intervals. \ref{fig:agg_bin_intervals_and_coverage}: \textbf{(Left)} Bin count intervals from one out of the $M$ samples using post-inversion aggregation with the Misspecified GMM Ansatz. \textbf{(Right)} Estimated bin-wise coverage of the post-inversion aggregation approach shows these intervals have the desired coverage.}
\end{figure}

Since the misspecification is exacerbated by the wide-binning, one strategy is simply to use more fine bins, mitigating bin-wise ansatz misspecification. Staying with $m=40$ smeared bins, with the least-squares intervals, we are restricted to at most $n=40$ true bins so that the smearing matrix retains full-column-rank. With more true bins, we obtain results as those shown in Fig.~\ref{fig:fine_bin_fix}. This figure shows that unfolding with more true bins nearly fixes the coverage problem shown in the right side of Fig.~\ref{fig:wide_bin_ls_failure}, but at the expense of creating significantly wider bin-wise intervals.
Ideally, we want to obtain intervals with widths like those on the left of Fig.~\ref{fig:wide_bin_ls_failure}, but with the coverages like those on the right of Fig.~\ref{fig:fine_bin_fix}. The intervals shown in Fig.~\ref{fig:wide_bin_ls_failure} were constructed by first discretizing the true space into ten bins and then finding the least-squares intervals directly. However, we could alternatively invert at the finest possible binning (as shown in Fig.~\ref{fig:fine_bin_fix}) subject to least-squares assumptions (namely, that $\bm{K}$ be full rank) and then aggregate the fine-bin intervals to the same coarse-bin level. We refer to this wide-bins-via-fine-bins strategy as ``post-inversion aggregation''. In the above example, this paradigm leads us to create a new matrix, $\bm{K} \in \mathbb{R}^{40 \times 40}$, corresponding to $n=40$ true bins which are then aggregated into the original 10 bins post-inversion using a sequence of functionals $\{ \bm{h} \}_{i = 1}^{10}$. Changing this order of operations yields significantly wider intervals, as seen in the left portion of Fig.~\ref{fig:agg_bin_intervals_and_coverage} compared with the intervals in the left portion of Fig.~\ref{fig:wide_bin_ls_failure}. However, in exchange for wider intervals, the right portion of Fig.~\ref{fig:agg_bin_intervals_and_coverage} shows that the coverage is now at the desired level across all bins. This happens because with $n=40$ true bins, the dependence of $\bm{K}$ on the MC ansatz is reduced and, as a result, the systematic misspecification in $\bm{K}$ is no longer large enough to cause a substantial bias in the unfolded solutions. In this particular example, as seen in the right portion of Fig.~\ref{fig:fine_bin_fix}, even using using a finer true binning does not entirely fix the coverage deficiency caused by the ansatz misspecification. As such, it is perhaps fortuitous that the post-inversion aggregation results in the right portion of Fig.~\ref{fig:agg_bin_intervals_and_coverage} show nominal coverage. However, the coverage improvement with the finer bins is clear and ideally, we would reduce the bin width further until we can be confident that the sensitivity to the ansatz misspecification is negligible. With the least-squares intervals, the full-rank constraint sets up a barrier for finer binning but the other methods presented herein do not have that limitation, as demonstrated in Section~\ref{sec:dealing_w_adv_ansatz}.

\subsection{Enforcing Non-negativity Improves Interval Width} \label{sec:add_non_neg_constraint}
With the bin-wise coverage now at nominal level, one might be tempted to conclude that these intervals are the best we can do. However, some of the intervals shown in the left portion of Fig.~\ref{fig:fine_bin_fix} used in the aggregations violate the known physical constraints of the unfolding problem; namely, that bin counts must be non-negative. Hence, constructing intervals containing negative values indicates that some key information is absent from the procedure.

\begin{figure}[htp]
    \centering
    \includegraphics[width=0.9\textwidth]{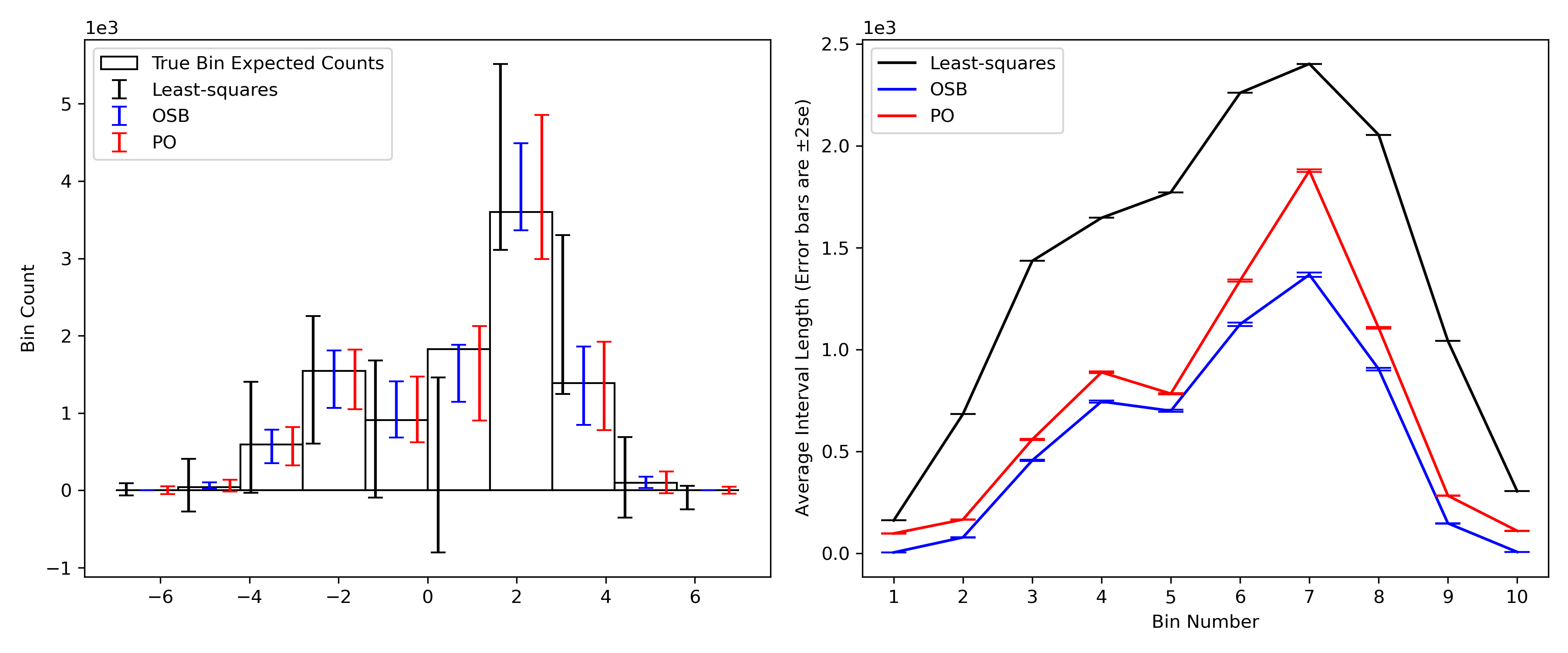}
    \caption{\textbf{(Left)} Comparing least-squares, OSB, and PO intervals for one realization of data shows that incorporating the non-negativity constraint dramatically reduces interval width. \textbf{(Right)} Expected interval width comparison between the least-squares, OSB, and PO interval constructions shows that incorporating the non-negativity physical constraint dramatically shortens the constructed confidence intervals. The error bars show the standard error of the average interval widths in order to demonstrate that the expected widths are significantly different. Additionally, the least-squares intervals are fixed width, so their standard error is zero.}
    \label{fig:expected_int_length_full_rank}
\end{figure}

The OSB and PO intervals are both capable of including this physical constraint. Using this additional information in the optimization has a clear benefit for the expected width of the constructed intervals. Indeed, the right portion of Fig.~\ref{fig:expected_int_length_full_rank} shows that the least-squares intervals are uniformly wider in expectation than both the OSB and PO intervals constructed with $\mathbf{A} = -\mathbf{I}_n$ and $\bm{b} = \bm{0}$ to enforce the non-negativity constraint. These expected widths are estimated using the same $M_D = 1{,}000$ samples used to estimate the coverage in the previous sections, with the error bars computed as twice the mean's sample standard error. More specifically, but less generally, the left portion of Fig.~\ref{fig:expected_int_length_full_rank} illustrates how intervals generated for one data sample are dramatically shortened when using the non-negativity constraint in the optimization. As presented in Section~\ref{sec:po_intervals}, the PO intervals require a prior expectation. To refrain from adding additional complexity here, the PO intervals constructed to make Fig.~\ref{fig:expected_int_length_full_rank} use a uniform prior, with bin counts set to the average true bin count, i.e.,
\begin{equation} \label{eq:flat_prior}
   \bm{m}_{\blambda} := \frac{\blambda^\top \mathbf{1}}{n} \cdot \mathbf{1}.
\end{equation}
We consider the effect of using different priors in Section~\ref{sec:further_simulations}, where we show that the PO intervals have little sensitivity to the choice of the prior.

In addition to providing shorter intervals, both the OSB and PO intervals maintain coverage guarantees as shown in Supplement Fig.~\ref{fig:coverage_guarantee_full_rank}. In fact, both the OSB (left) and PO (right) intervals over-cover relative to the desired confidence level on most bins. However, we note the nearly nominal coverage of the OSB intervals on the boundary bins shown in the left portion of Supplement Fig.~\ref{fig:coverage_guarantee_full_rank}. Given the nature of the true intensity, these bins lie on the portions of the domain on which the intensity is nearly zero. Similar to an observation made in \cite{kuusela_stark}, it appears that having the true intensity close to the non-negativity constraint can produce this type of behavior. While over-coverage may indicate a lack of efficiency in the form of slack in the interval widths, it is clear that both the OSB and PO interval widths are good relative to the least-squares~intervals.
\subsection{Handling an Adversarial Ansatz} \label{sec:dealing_w_adv_ansatz}
If systematic error from the ansatz can cause coverage problems in the wide-bin unfolding setting as shown in Section \ref{sec:wide_bin_problem}, it may be the case that there exists an ansatz which induces enough systematic error to even break the coverage shown for the OSB and PO intervals in Supplement Fig.~\ref{fig:coverage_guarantee_full_rank}. Exploring this scenario was the motivation for creating the Adversarial Ansatz as shown in Fig.~\ref{fig:intensity_functions}. 

\begin{figure}[htp]
    \centering
    \includegraphics[width=0.85\textwidth]{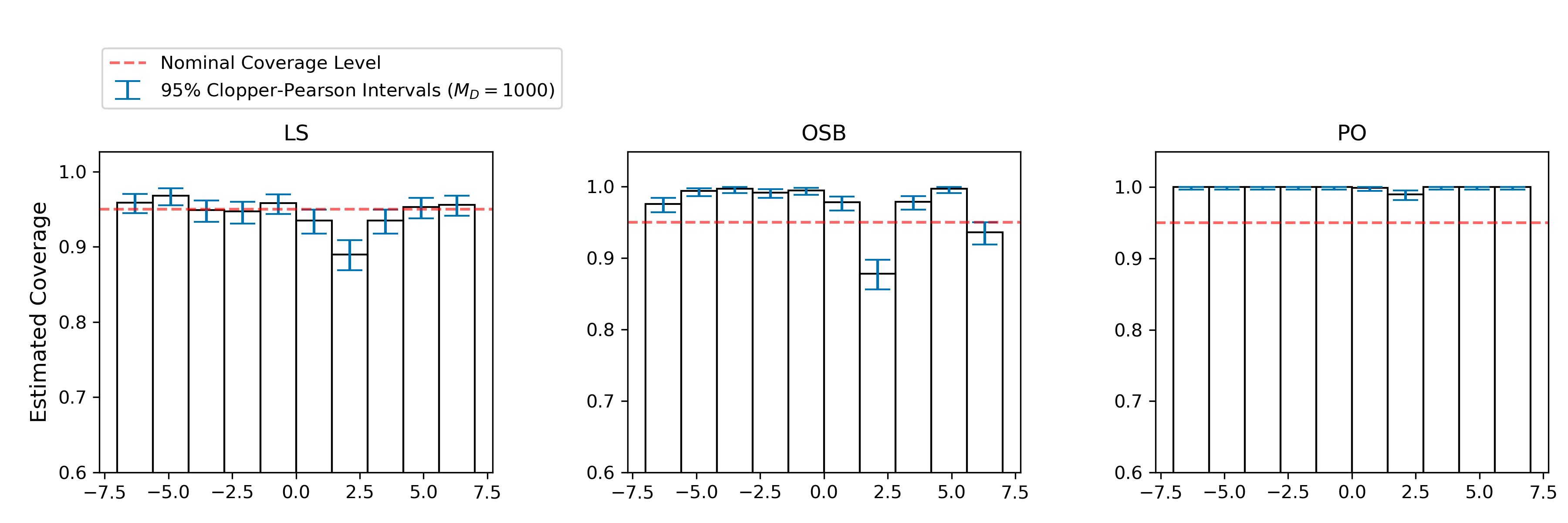}
    \caption{Coverage of (from left to right) least-squares, OSB, and PO 95\% intervals using a smearing matrix constructed with the Adversarial Ansatz. This ansatz leads to systematic error in the smearing matrix, creating severe lack of empirical coverage for both the least-squares and OSB intervals.}
    \label{fig:adversarial_ansatz_coverage_95percent}
\end{figure}

\begin{figure}[htp]
    \centering
    \includegraphics[width=0.7\textwidth]{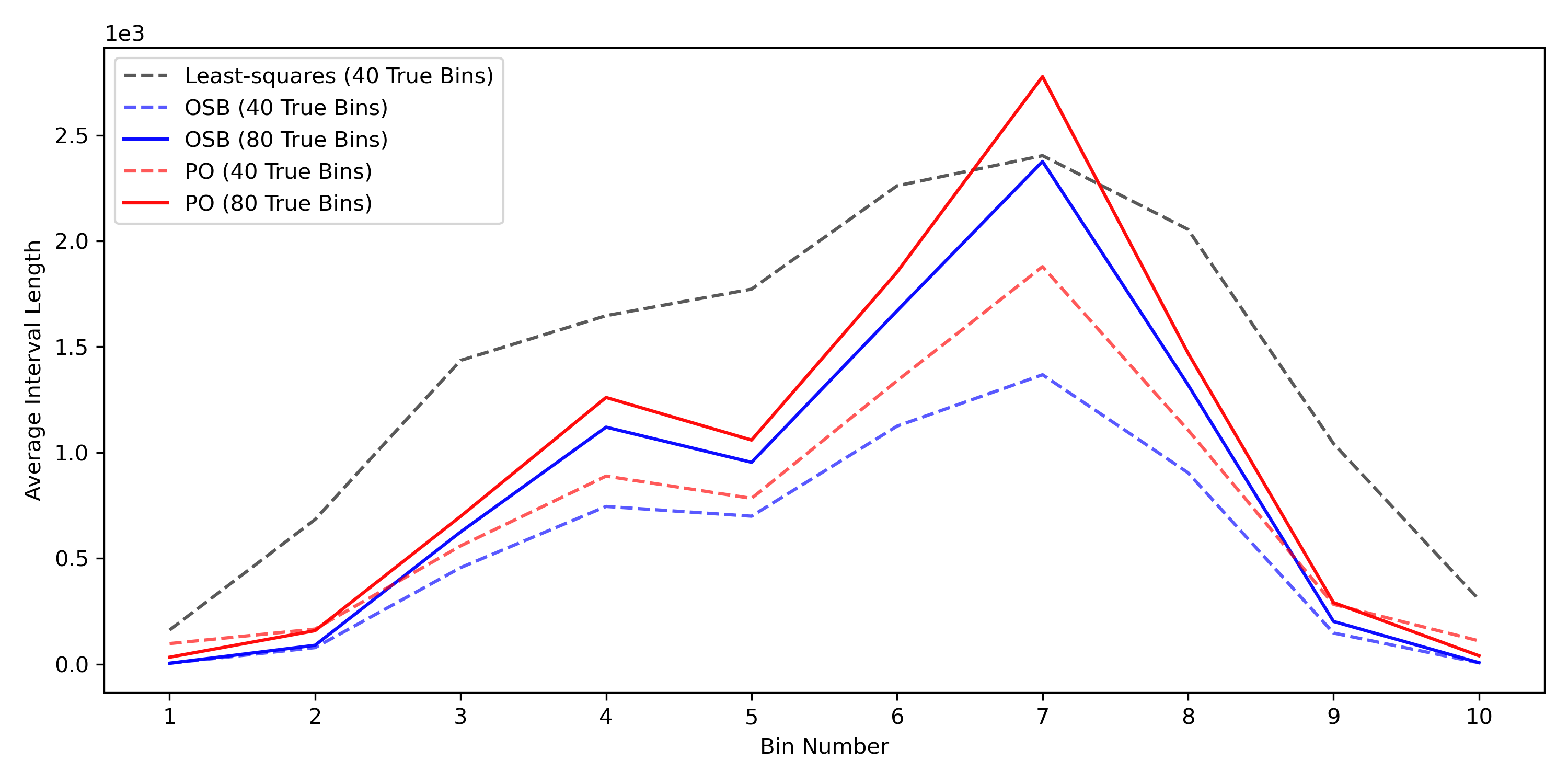}
    \caption{Expected 95\% interval widths when using $\bm{K} \in \mathbb{R}^{40 \times 80}$ and more true bins than in the original full-rank smearing matrix. For almost all bins, for both the OSB and PO intervals, the 80 true bin expected interval width is greater than that of the 40 true bin configuration.}
    \label{fig:adversarial_80bin_expected_lengths}
\end{figure}

To explore this potential failure mode, we compute a smearing matrix $\bm{K} \in \mathbb{R}^{40 \times 40}$ with $n=40$ true bins, as in Section \ref{sec:addressing_wide_bin_problem}, but this time using the Adversarial Ansatz for $f^{\textrm{MC}}$, and estimate the coverage of 95\% intervals in the same manner as above for the least-squares, OSB, and PO procedures. The results of the 95\% interval estimation in Fig.~\ref{fig:adversarial_ansatz_coverage_95percent} show the coverage of least-squares, OSB, and PO intervals from left to right. Since we have already demonstrated the effect of the wide-bin bias on the least-squares intervals, it is not surprising that a few of those intervals show under-coverage with the Adversarial Ansatz despite using a large number of true bins. But, here the systematic misspecification is large enough to also affect the coverage of the OSB intervals as shown by their significant under-coverage in the seventh bin.
One solution is to further circumvent the systematic error induced by the Adversarial Ansatz by using an even finer binning in the true space during inversion and then adjusting each bin functional to aggregate more of the fine bins to arrive at the same ten final bins. Assuming that we keep the number of smeared bins fixed at $m=40$, we end up with a column-rank-deficient smearing matrix, and therefore cannot use the least-squares intervals. However, there is no such full-rank requirement for the OSB and PO intervals. We therefore construct a new smearing matrix, $\bm{K} \in \mathbb{R}^{40 \times 80}$ using $n=80$ true bins, and perform the same coverage estimation as above for 95\% intervals. The results of this experiment are shown in Supplement Fig.~\ref{fig:adversarial_ansatz_coverage_68percent_fix}. Now, both the OSB and PO intervals (left to right in Supplement Fig.~\ref{fig:adversarial_ansatz_coverage_68percent_fix}) have at least nominal coverage across all bins. By moving to the rank-deficient scenario, we were able to use enough true bins to reduce the systematic misspecification in $\bm{K}$ to a level where the OSB intervals are no longer affected by the wide-bin bias.
Though using a finer true binning fixes the coverage issues shown in Fig.~\ref{fig:adversarial_ansatz_coverage_95percent} and gives increased protection against misspecification of $f^\text{MC}$, we pay by increasing the expected interval widths as shown in Fig.~\ref{fig:adversarial_80bin_expected_lengths}. Both the PO the OSB intervals experience substantial width inflation in the rank-deficient regime relative to the full-rank scenario. Nevertheless, their expected widths are almost uniformly shorter than those of the full-rank least-squares intervals, which are both longer and undercover in this scenario. Since OSB's seventh bin exhibits under-coverage in Fig.~\ref{fig:adversarial_ansatz_coverage_95percent}, the increase in interval width is reasonable. It may be noted that the expected least-squares interval width for bin seven is less than the interval width for the 80-bin PO interval. However, the PO intervals have coverage while the least-squares intervals do not so we would still opt for the PO intervals among these two.

In summary, we have shown that with the Adversarial Ansatz, both the OSB and PO intervals can be constructed to provide coverage and, in most cases, out-perform the width of the least-squares intervals, which either undercover or are not applicable in this scenario. This was enabled by the ability of the OSB and PO intervals to make use of the physical constraints and to handle a rank-deficient $\bm{K}$ while still maintaining frequentist coverage.


\subsection{Further Simulations} \label{sec:further_simulations}
The sections above provide some insight into the OSB and PO intervals' ability to incorporate physical constraints and to address the coverage issues in least-squares intervals that arise from wide-binning due to the systematic error in the MC ansatz. In this section, we provide additional interval method comparisons against the OSB and PO intervals, demonstrate that using more bins in the true space does not cause interval widths to diverge, and provide evidence that the expected PO interval widths are robust to the choice of the prior.

\subsubsection{Comparison against simultaneous strict bounds and minimax intervals}

\begin{figure}[htp]
    \centering
    \includegraphics[width=0.7\textwidth]{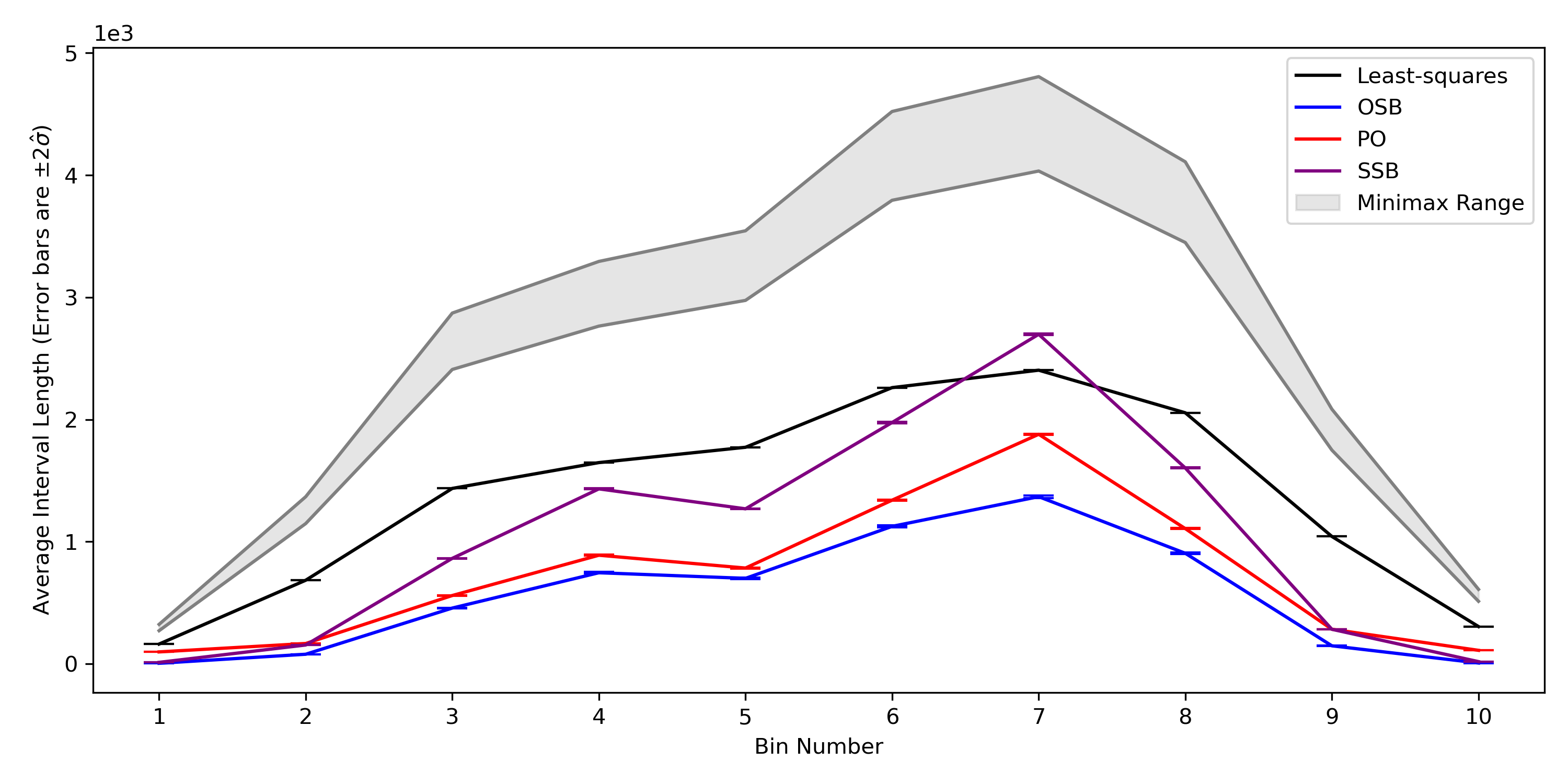}
    \caption{Expected 95\% interval widths for LS, OSB, PO, and SSB intervals, shown with upper and lower bounds for the minimax interval widths. All intervals are constructed with the full-rank smearing matrix. Since the minimax intervals are the most conservative, the least-squares intervals do not take physical constraints into account, and the SSB intervals are simultaneous intervals applied one at a time, it makes sense the PO and OSB intervals are uniformly narrower across almost all bins. Between the OSB and PO intervals, the OSB intervals are narrower since each interval optimizes width with respect to the observed data, whereas the PO intervals optimize width offline.}
    \label{fig:expected_widths_across_methods}
\end{figure}

In the above analysis (e.g., as seen in Fig.~\ref{fig:expected_int_length_full_rank}), we compared the expected widths of the OSB and PO intervals constructed using the full-rank smearing matrix and the Misspecified GMM Ansatz with the fixed-width least-squares intervals. The width improvement for the OSB and PO intervals over the least-squares intervals is expected since the latter do not incorporate the non-negativity constraint. In order to compare against other methods that also account for constraints, we consider the expected interval widths of the SSB intervals and minimax intervals, as described in Section~\ref{sec:other_intervals}. We perform the same procedure as described in Section~\ref{sec:add_non_neg_constraint} in order to estimate the expected width of the SSB intervals. Like the least-squares intervals, the minimax intervals are fixed-width, and hence have interval width that is independent of any particular realization of data (although the actual width is only known up to a fixed range, as described in Section~\ref{sec:other_intervals}). The results of these simulations are shown in Fig.~\ref{fig:expected_widths_across_methods}.

Since we are able to only find a lower and upper bound for the minimax interval widths, we shade the region between these bounds to indicate where the actual minimax interval widths would be for each bin. We observe that the OSB and PO intervals still have the shortest expected width for almost all bins (excluding the end bins) when compared against the other three alternatives. This result is sensible since the SSB intervals are simultaneous intervals, and are therefore conservative if evaluated as one-at-a-time intervals. Furthermore, the minimax intervals are, by definition, the most conservative fixed-width affine intervals for this setup, aligning with the observation that they are uniformly the widest of these~intervals.

\subsubsection{Interval widths as a function of the number of true bins}

\begin{figure}
    \centering
    \begin{subfigure}[b]{0.49\textwidth}
    	\includegraphics[width=\textwidth]{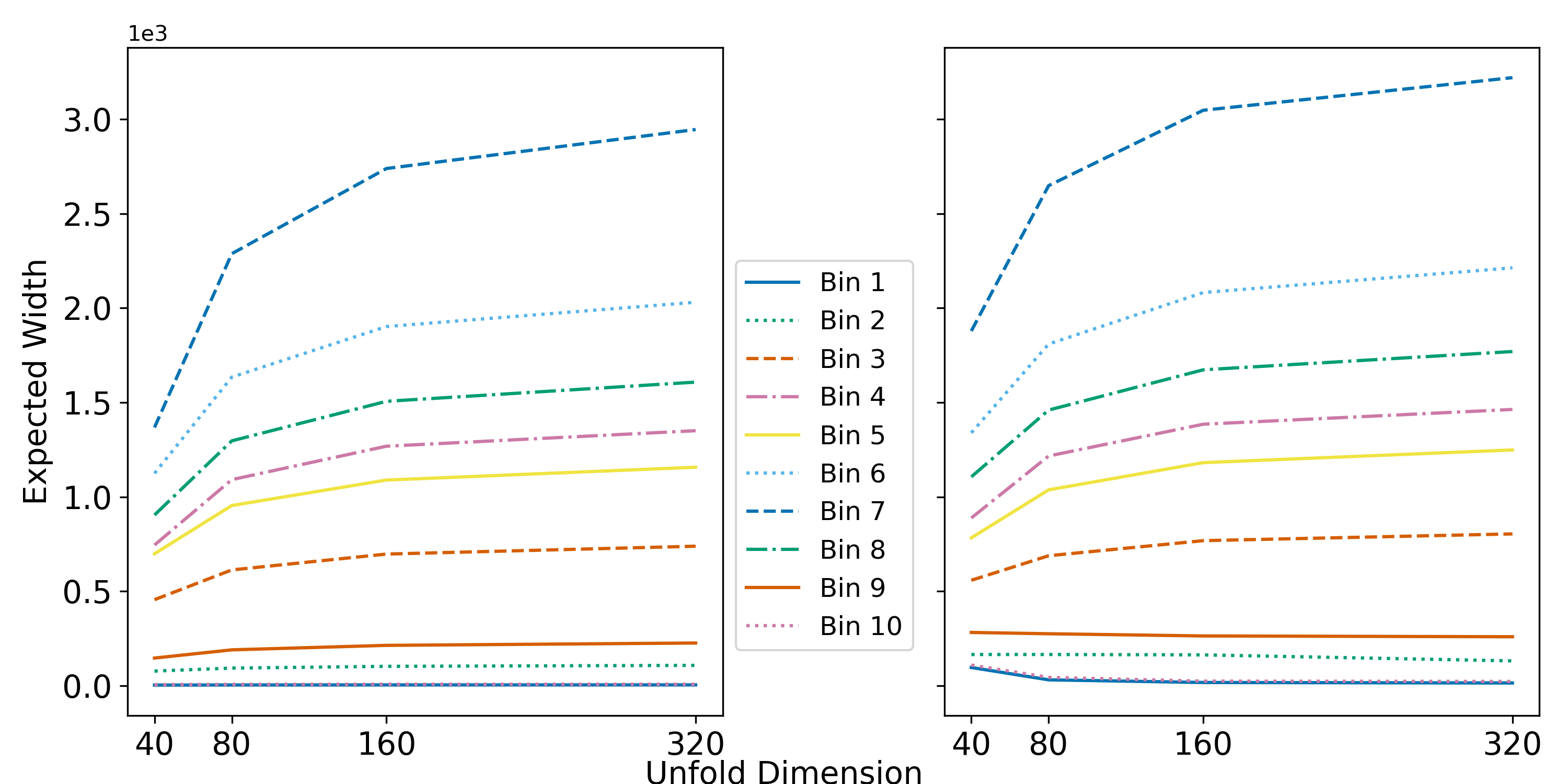}
	\caption{}
    	\label{fig:expected_widths_across_unfold_dim}
    \end{subfigure}
    \begin{subfigure}[b]{0.49\textwidth}
    	\includegraphics[width=\textwidth]{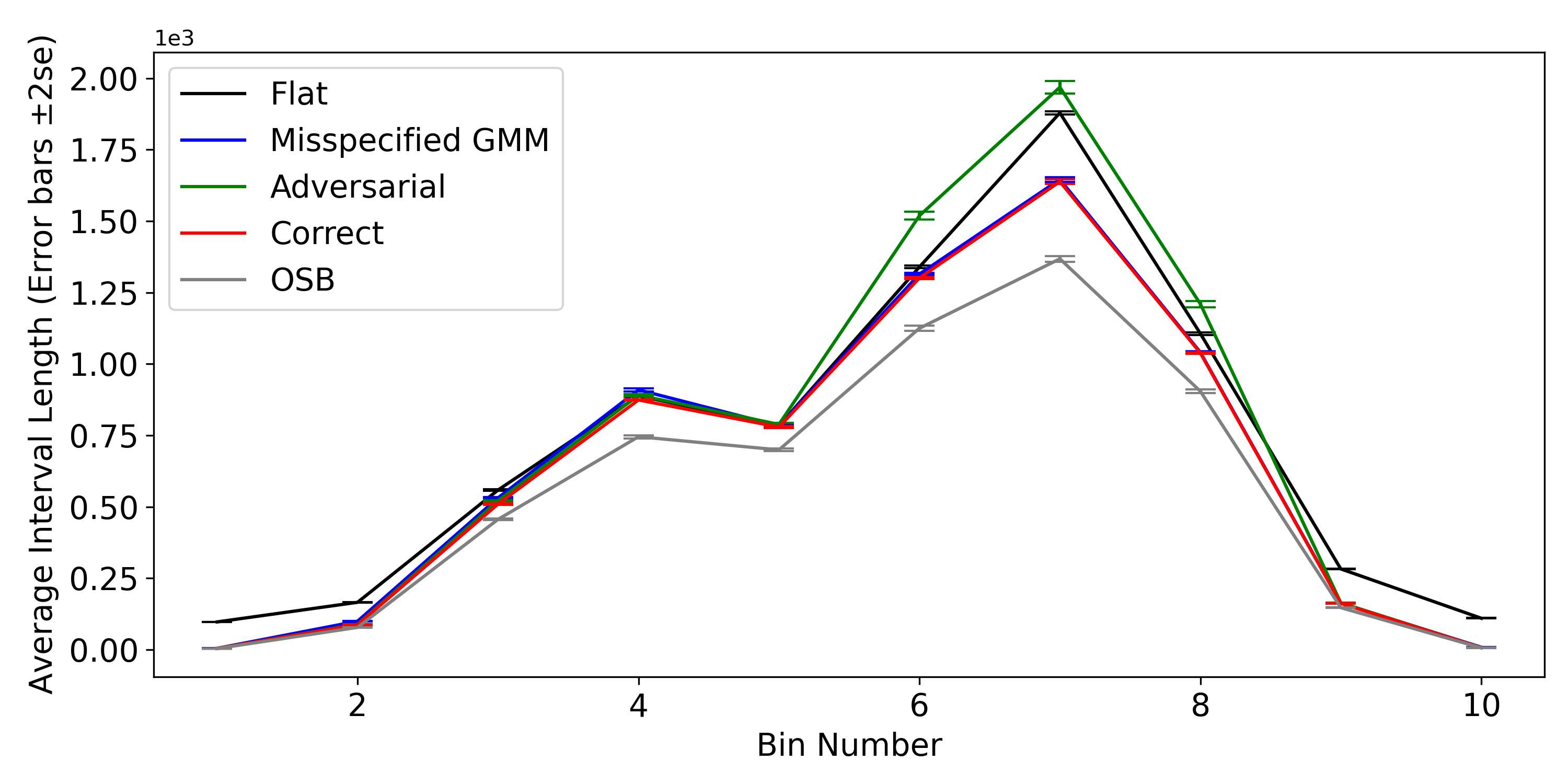}
	\caption{}
    	\label{fig:prior_choice_and_expected_length}
    \end{subfigure}
    \caption{\ref{fig:expected_widths_across_unfold_dim}: Estimated expected 95\% interval widths as a function of the number of true unfolding bins provide assurance that both OSB and PO intervals remain finite as the number of true bins increases. Both OSB \textbf{(Left)} and PO \textbf{(Right)} interval widths become less sensitive to unfolding dimension as the number of true bins increases. \ref{fig:prior_choice_and_expected_length}: The choice of prior used when optimizing 95\% PO interval does not have a substantial impact on the expected interval widths. We observe that for three incorrectly specified priors (Flat, Misspecified GMM, and Adversarial), in all but three bins, the expected interval widths are close. In bins 6, 7, and 8, we observe the width benefits of having a more correctly specified prior. We also observe the closeness of most bin expected interval widths to the expected OSB interval widths.}
\end{figure}

One of the primary practical benefits of the OSB and PO intervals is that they can be constructed even with rank-deficient smearing matrices. As we demonstrated above by increasing the number of true bins from 40 to 80, this flexibility can be used to overcome the systematic error that exists due to the MC ansatz. However, this ability is questionable if the width of the intervals diverges as a function of the number of true bins. We know from previous work \cite{kuusela_stark} that the SSB intervals are finite even for infinite-dimensional true spaces. Since we expect the OSB and PO intervals to be shorter than the SSB intervals, we expect these intervals to remain finite as the number of true bins is increased. To provide assurance that the interval widths do not diverge, we repeat the simulation study in Section~\ref{sec:dealing_w_adv_ansatz}, but additionally compute the OSB and PO intervals using 160 and 320 true bins in addition to the 40- and 80-bin setups. For this study, we use the Misspecified GMM Ansatz to construct each smearing matrix and a flat prior (see Eq.~\eqref{eq:flat_prior}) for the PO intervals. The results of these experiments can be seen in Fig.~\ref{fig:expected_widths_across_unfold_dim} showing the expected width as a function of the binning setup, with one line for each aggregated bin. Across all bins, and for both OSB and PO intervals (left and right, respectively), we observe that the interval width stabilizes as a function of the number of true bins. As such, if we need to increase the number of true bins to circumvent the wide-bin systematic error, we can be reasonably sure that the interval widths will not diverge.

\subsubsection{Robustness against the choice of the prior}

Not only are the PO interval expected widths robust to the binning setup, but they also exhibit a degree robustness with respect to the choice of the prior. We explore this robustness by repeating the simulation in Section~\ref{sec:add_non_neg_constraint} with the $40\times40$ full-rank smearing matrix. In addition to constructing the PO intervals with a flat prior (see Eq.~\eqref{eq:flat_prior}), we consider three additional priors. First, we compute the bin means of the true data generating function (see Eq.~\eqref{eq:simulation_intensity}) to compute the \emph{Correctly Specified Prior}, followed by the bin means of the Misspecified GMM Ansatz to create the \emph{Misspecified GMM Prior}, and finally the Adversarial Ansatz to create the \emph{Adversarial Prior}. The expected interval widths for each bin for each prior can be seen in Fig.~\ref{fig:prior_choice_and_expected_length}. The Adversarial Prior, the most misspecified out of the four, shows the largest expected width in most of the bins. The Misspecified GMM Prior creates intervals with expected widths close to the correctly specified prior. This result is sensible since the Misspecified GMM is close to the true data generating process. For half of the bins, the choice of the prior does not appear to lead to substantively different results, but in some bins (especially bins 6, 7, and 8), having a more correctly specified prior appears to shorten the interval width. Notably, using the Misspecified GMM Prior (or the Correctly Specified Prior) would make the PO interval widths comparable to the OSB intervals.

\section{Application to Unfolding a Steeply Falling Particle Spectrum} \label{sec:particle_spectrum}
While the above results in Section \ref{sec:story_section} demonstrate the properties of the OSB and PO intervals in a wide-bin deconvolution setting, the data generating process was not directly motivated by a specific particle physics data analysis scenario. In this section, we use the inclusive jet transverse momentum spectrum \cite{cms_2011} as a more concrete unfolding problem in particle physics. This spectrum reflects the production rate of jets (collimated streams of particles) as a function of the jet (transverse) momentum at proton-proton collisions at the Large Hadron Collider at CERN. The associated intensity is an example of a steeply falling particle spectrum, for which the intensity rapidly decays for larger transverse momentum ($p_T$) values. We follow the test setup equations and parameters outlined in \cite{kuusela_phd_thesis} (see Section 3.4.2). In the same way as we defined a true and ansatz intensity function for the deconvolution example in Section \ref{sec:story_section}, we define a true intensity using the parameters in Section 3.4.2 of \cite{kuusela_phd_thesis}, and an ansatz intensity using the alternative parameters in Section 4.2 of \cite{kuusela_phd_thesis}. Supplement Fig.~\ref{fig:steeply_falling_spectrum_intensity} shows both intensities in the left panel and the fine-bin and wide-bin discretized true intensity functions in the middle and right panels.
As seen in Supplement Fig.~\ref{fig:steeply_falling_spectrum_intensity}, we consider the intensity function from $400$ to~$1{,}000$ GeV.

Like the wide-bin deconvolution problem, we find confidence intervals for $10$ wide bins in two ways. First, we provide a baseline by directly computing the wide-bin intervals using least-squares via a smearing matrix built on $10$ true and $30$ smeared bins. Second, we compute the OSB, PO and SSB intervals with a rank-deficient smearing matrix built on $60$ true and $30$ smeared bins to again reduce the magnitude of the systematic error induced by the misspecified ansatz intensity. To get a sense of the misspecification in this problem, consider Supplement Fig.~\ref{fig:matrix_misspec}. This figure shows $\lvert K_{ij} - K^{\textrm{MC}}_{ij} \rvert$ for all rows $i$ and columns $j$, plotted on a logarithmic scale. The left panel in Supplement Fig.~\ref{fig:matrix_misspec} shows the difference between the true and ansatz smearing matrices for the $30 \times 10$ case, i.e., the case in which intervals are computed directly on the wide bins. The right panel in Supplement Fig.~\ref{fig:matrix_misspec} show the same difference but for the $30 \times 60$ matrix, i.e., the rank-deficient case. The misspecification in the $30 \times 10$ case is $\mathcal{O}(10^{2})$ times larger than the misspecification in the $30 \times 60$ case.

\begin{figure}[htp]
    \centering
    \includegraphics[width=0.7\textwidth]{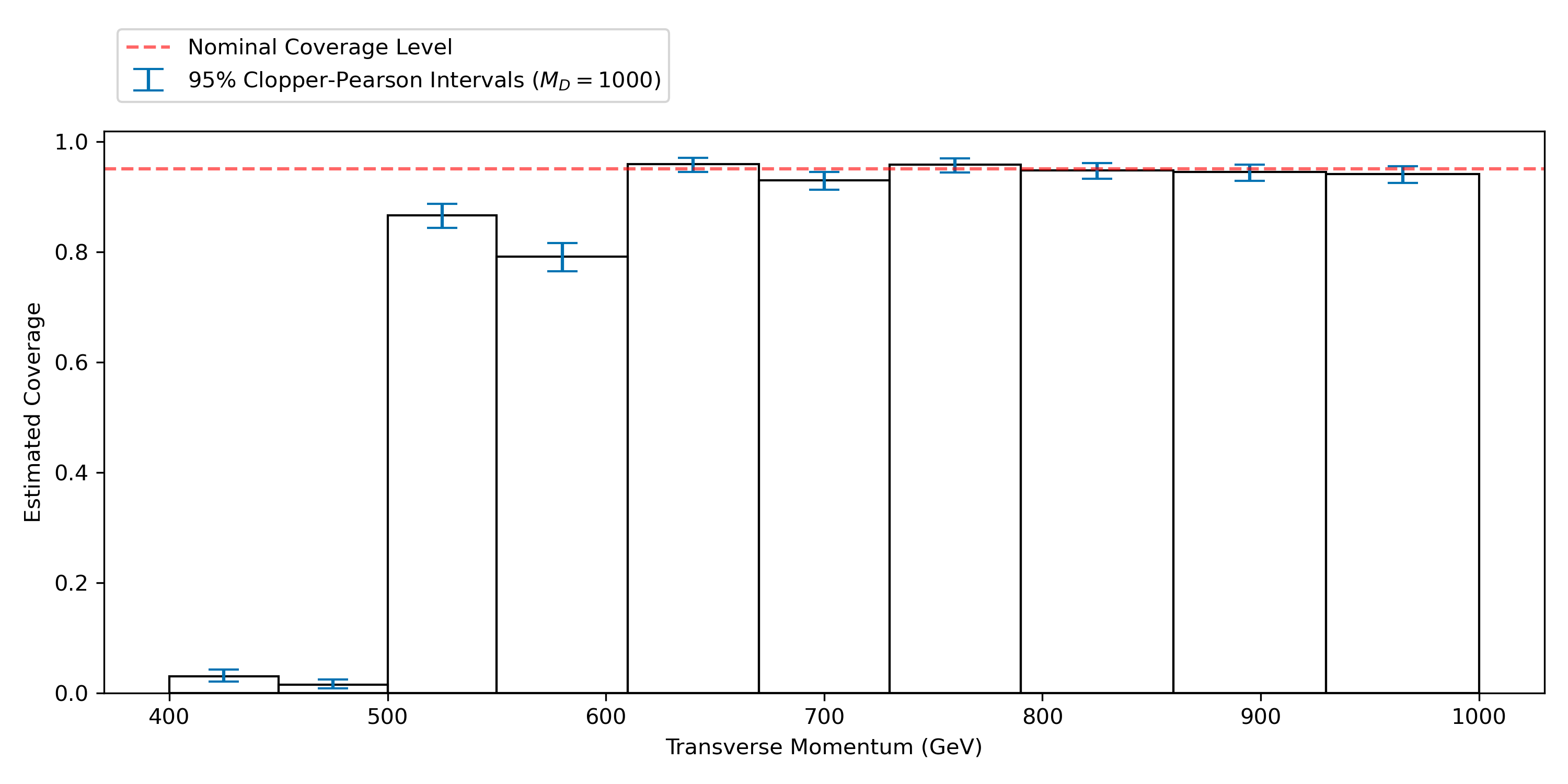}
    \caption{Wide-bin unfolding in the steeply falling spectrum example with least-squares intervals produces dramatic undercoverage in several bins.}
    \label{fig:steep_fall_ls_ints}
\end{figure}

\begin{figure}[htp]
    \centering
    \includegraphics[width=0.8\textwidth]{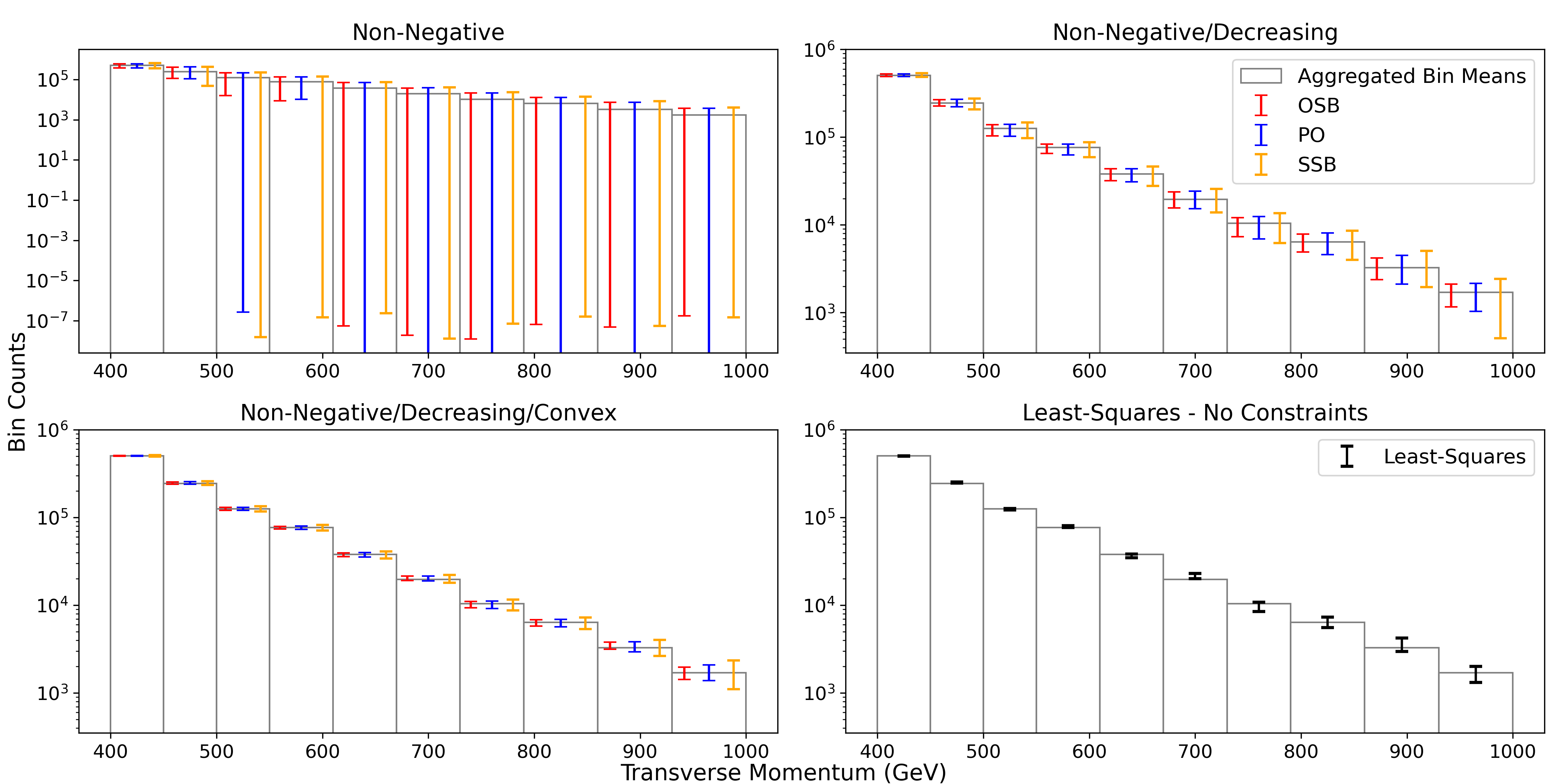}
    \caption{Example 95\% wide-bin intervals across different procedures and constraint configurations for the steeply falling particle spectrum based on a $30 \times 60$ smearing matrix. Adding shape constraints significantly shortens interval widths across all procedures. In the upper left plot, the PO intervals intersecting the horizontal axis have a lower bound close to zero. The top two and bottom left plots show OSB intervals as the shortest, PO as the middle width, and SSB intervals as the widest, across most bins. Example least-squares intervals are included in the bottom right and essentially show interval widths comparable to the OSB/PO/SSB intervals in the bottom left panel, but as shown in Fig.~\ref{fig:steep_fall_ls_ints} the wide-bin least-squares intervals do not have correct coverage.}
    \label{fig:sample_intervals}
\end{figure}

In the wide-bin deconvolution example, directly computing wide-bin confidence intervals via least squares produces intervals lacking coverage because of the systematic error induced by the misspecified ansatz. As shown in Fig.~\ref{fig:steep_fall_ls_ints}, least-squares intervals computed in this scenario similarly lack coverage. Further, the bottom right panel in Fig.~\ref{fig:sample_intervals} shows the wide-bin least-squares intervals for one realization of data. As expected, these intervals are very narrow on the bins where they lack coverage, namely, the first four bins, making them sensitive to the wide-bin bias induced by the misspecification of the ansatz.
We now proceed to show that the coverage problems displayed in Fig.~\ref{fig:steep_fall_ls_ints} can again be ameliorated by using the rank-deficient setup to form the wide-bins-via-fine-bins intervals using the OSB, PO or SSB methods. Although we use $10$ wide bins in this example as we did in the deconvolution example, reference \cite{kuusela_phd_thesis} shows in Eq.~(3.36) that the additive noise used to model the smearing is heteroskedastic as a function of the $p_T$, meaning that the wide-bin width should ideally vary across the $p_T$ domain, unlike the constant wide-bin width in the homoskedastic deconvolution example. The physical motivation for this choice is that the wide-bin widths should be comparable to the resolution of the measurement apparatus, which in this case varies as a function of~$p_T$. For the range of $p_T$ considered, the noise standard deviation is characterized by $\sigma(p_T) \propto \sqrt{p_T}$. As such, the wide-bin widths in $p_T$ are enlarged at the same proportional rate. Namely, for wide-bin width $B$, as $p_T$ increases, $B$ is increased such that $B \propto \sqrt{p_T}$. The resulting endpoints of these bins are then matched to the closest endpoints of the uniformly sized $60$ true fine bins. As a result, Supplement Fig.~\ref{fig:steeply_falling_spectrum_intensity} shows that the left-most wide bin includes five fine bins while the right-most includes seven fine bins.

In addition to the inclusive jet transverse momentum spectrum providing a more realistic example of unfolding than the simple deconvolution setup, as explained in \cite{kuusela_stark}, there are physically motivated intensity function shape constraints that can be included when optimizing the intervals in addition to the non-negativity constraint used in the deconvolution example. Namely, we expect the intensity to be monotonically decreasing and convex. These constraints are implemented via the $\mathbf{A}$ matrix as seen in the optimization problem~\eqref{opt:osb_lower}, for example.

When optimizing over the parameter $\blambda \in \mathbb{R}^n$, the non-negativity, monotonically decreasing, and convexity shape constraints can be implemented as follows. For the infinite-dimensional version of implementing these shape constraints, see \cite{kuusela_stark}. The non-negativity implementation is already explained in Section~\ref{sec:osb_description}. Denote this constraint matrix by $\mathbf{A}^n \in \mathbb{R}^{n \times n}$. To implement the decreasing constraint, we first note that for all $i \in [n - 1]$, $\lambda_i \geq \lambda_{i + 1}$ must hold. When working with the optimizations like program~\eqref{opt:osb_lower}, these necessary conditions can be met by constructing $\mathbf{A}^d \in \mathbb{R}^{(n - 1) \times n}$ such that $\mathbf{A}^d_{i,i} = -1$ and $\mathbf{A}^d_{i,i + 1} = 1$, for all $i \in [n - 1]$. To create a necessary convexity condition, we observe that for three adjacent elements, e.g., $\lambda_{i}$, $\lambda_{i + 1}$ and $\lambda_{i + 2}$, in order for the vector $\blambda$ to follow a convex shape, it must be true that for all $i \in [n - 2]$, we have $\frac{\lambda_i + \lambda_{i + 2}}{2} \geq \lambda_{i + 1}$. Thus, we must have the following inequality for all $i \in [n - 2]$: $-\lambda_{i} + 2 \lambda_{i + 1} - \lambda_{i + 2} \leq 0$. These necessary conditions can be met by creating a matrix $\mathbf{A}^c \in \mathbb{R}^{(n - 2) \times n}$ such that $\mathbf{A}^c_{i,i} = -1$, $\mathbf{A}^c_{i,i + 1} = 2$, and $\mathbf{A}^c_{i,i + 2} = -1$, for all $i \in [n - 2]$. Combining these constraints is accomplished simply by stacking these individual shape constraint matrices on top of each other. In particular, we work with the same three constraint setups as those shown in Table 1 of \cite{kuusela_stark}, namely, non-negativity; non-negativity and monotonically decreasing; and non-negativity, monotonically decreasing, and convex. These setups are referred to as N, ND, and NDC, respectively. Here we assume that the fine bins have a uniform width. If the fine-bin discretization was done using variable bin widths, the matrices $\mathbf{A}^d$ and $\mathbf{A}^c$ would need to be adjusted to account for the variable bin widths.

Like the deconvolution example, we are primarily interested in evaluating the coverage and expected width of the intervals created with each of the above three constraint setups. We estimate the coverage and interval width by sampling $M_D = 1{,}000$ draws from the distribution shown in Eq.~\eqref{eq:true_poisson_data_gen}, and the smearing matrix is constructed as described above. In particular, we build OSB, PO, and SSB intervals with each of the above three constraint combinations, for a total of nine different interval setups. First, Fig.~\ref{fig:sample_intervals} shows one realization of the $1{,}000$ intervals constructed for each interval procedure and for each aggregated bin. Despite the rapid interval width decay as a function of the number of shape constraints, Supplement Fig.~\ref{fig:steeply_falling_spectrum_coverage_full_rank} in Supplement Section~\ref{app:extra_plots} shows that across all interval procedures, constraint configurations, and aggregated bins, we retain at least nominal coverage, as desired.

\begin{figure}[htp]
    \centering
    \includegraphics[width=0.7\textwidth]{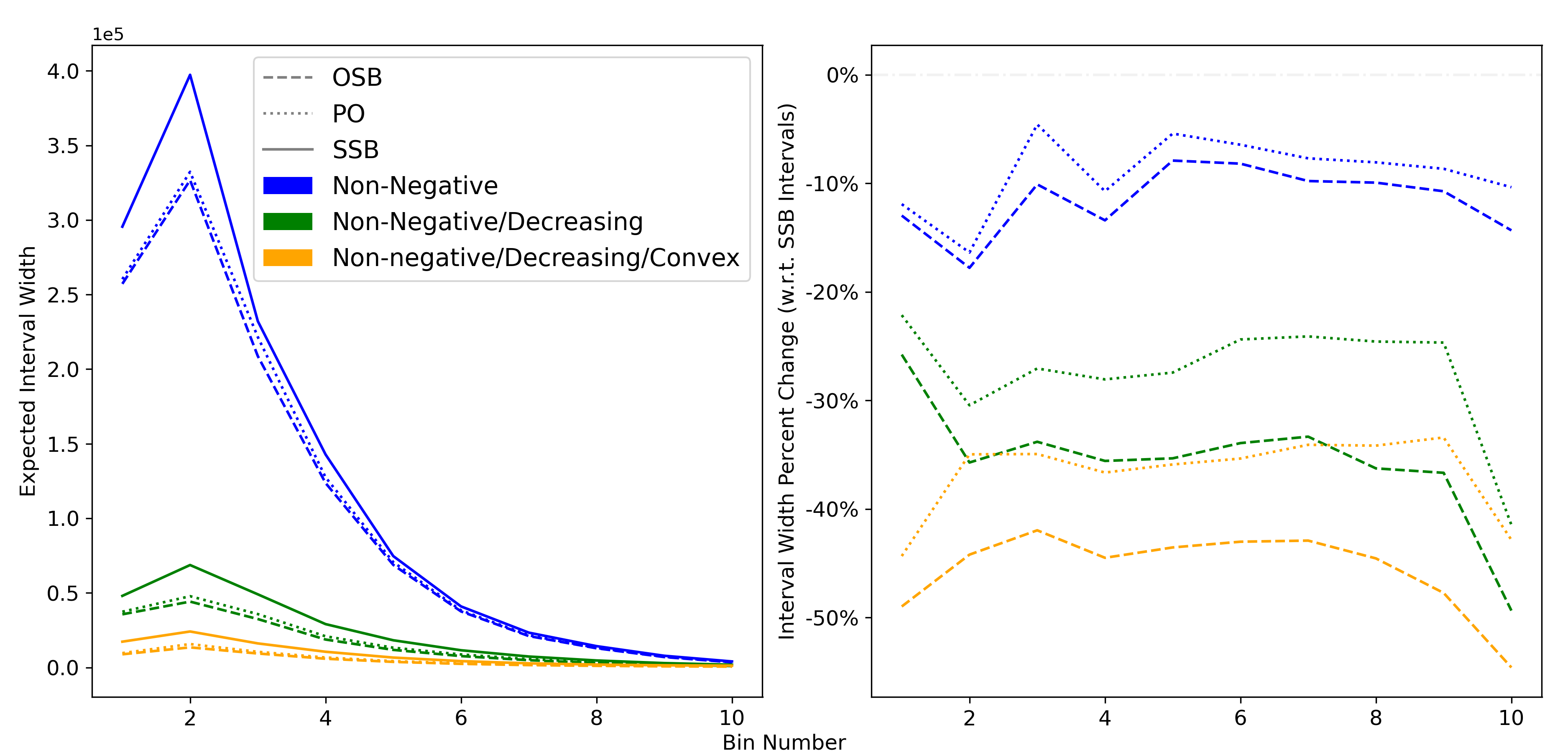}
    \caption{\textbf{(Left)} Expected 95\% interval widths across bins. Similar to the deconvolution example, the OSB intervals are uniformly shorter than the PO intervals, which are uniformly shorter than the SSB intervals. \textbf{(Right)} Percent decrease in expected interval width with respect to the SSB interval widths. Both the OSB and PO interval procedures produce shorter intervals across all bins and constraint configurations than SSB. The width improvement with respect to the SSB intervals increases as more shape constraints are added for both OSB and PO intervals.}
    \label{fig:osb_po_stark_full_rank_sfs_exp_len_comp}
\end{figure}

Interval widths decreasing as a function of the number of shape constraints also holds in expectation, as shown in Fig.~\ref{fig:osb_po_stark_full_rank_sfs_exp_len_comp}. Additionally, the left panel in Fig.~\ref{fig:osb_po_stark_full_rank_sfs_exp_len_comp} shows the same expected width ordering across interval procedures as previously in Section~\ref{sec:story_section}. Namely, OSB intervals are uniformly shorter in expectation than PO intervals, which are in turn uniformly shorter in expectation than the SSB intervals. For a few bins in higher $p_T$ values, these improvements are difficult to see. The improvements can be more easily observed in the right panel of Fig.~\ref{fig:osb_po_stark_full_rank_sfs_exp_len_comp}, showing the percent reduction in expected interval width using the SSB intervals as a baseline. This figure additionally shows that the percent reduction increases as shape constraints are added. For instance, the PO intervals are shorter than the SSB intervals for all bins and constraint configurations, but for the ND and NDC constraints, the PO intervals provide a much larger width improvement over the SSB intervals compared to the improvement with just the N constraint. The same applies to the OSB intervals as well.

This example provides a more concrete demonstration of how classical wide-bin unfolding with systematic error in the smearing matrix can nullify the typical coverage guarantees of least-squares intervals. This can be addressed using the wide-bins-via-fine-bins approach based on the OSB, PO and SSB intervals which are able to handle the nontrivial null space of the smearing matrix and incorporate additional physical constraints, while maintaining the desired coverage. It additionally demonstrates that the non-simultaneous OSB and PO methods offer substantial interval width improvements over the comparable SSB intervals, assuming that one-at-a-time coverage is what we are interested in.

\section{Discussion and Conclusions} \label{sec:discussion_conclusion}
When performing unfolding in high energy physics, wide-bin unfolding is an increasingly popular alternative for regularization. It reasonably sets wide bin widths to match the detector resolution. However, as we showed in Section~\ref{sec:story_section}, even a slight systematic error caused by the ansatz used to create the wide-bin smearing matrix can scupper the coverage guarantees of classical least-squares intervals. In this paper, we have proposed and substantiated a different approach to avoid these coverage issued in the aforementioned setup. Namely, instead of unfolding directly to the detector resolution, we propose a general methodology in which one first unfolds with as many true bins as possible and then aggregates these narrow-bin inversion results to the detector resolution. In order to maximally mitigate the systematic error caused by the ansatz misspecification, we invert with a rank-deficient matrix that has more true bins than smeared bins facilitated by the OSB and PO intervals.

In Sections~\ref{sec:story_section}~and~\ref{sec:particle_spectrum}, we explored the coverage and width properties of the OSB and PO intervals for a histogram deconvolution problem and for unfolding a steeply falling particle spectrum, respectively. In the deconvolution problem, the OSB and PO intervals clearly address the coverage deficiencies of the wide-bin least-squares intervals, while also providing superior interval width properties when compared against least-squares, SSB, and minimax intervals. Additionally, we demonstrated a low interval width sensitivity to the number of true bins (i.e., the extent to which the smearing matrix is rank-deficient) for both the OSB and PO intervals and to the choice of the prior for the PO intervals. Similarly, in the steeply falling spectrum example, we showed that the coverage and expected width properties of the OSB and PO intervals carry over to this more realistic unfolding setting from the simpler deconvolution example. This example further demonstrates the capacity and utility of these intervals to include shape constraints (i.e., non-negativity, monotonicity, and convexity) to further reduce the width of the optimized intervals while preserving coverage.

The OSB and PO intervals have properties that make them good methodological choices for the types of ill-posed inverse problems considered herein. However, determination of the ``best'' interval is largely dependent upon the specific application. As such, Table~\ref{table:comparison} provides an overview of some key properties of various intervals relevant to the statistical context of this paper. Note in Table~\ref{table:comparison} that the PO intervals are the narrowest intervals having empirical coverage, provable coverage, incorporation of physical constraints, and handling of rank-deficient linear models, which stem in part from their methodological novelty. In particular, the decision-theoretic framing allows the definition of a set of decision rules for which coverage is guaranteed. With this definition, we are able to focus the loss function only on interval width, as opposed to considering both width and coverage as in many other decision-theoretic treatments of confidence sets. Indeed, in previous literature \cite{evans_hansen_stark, evans_stark, schafer_stark, winkler}, decision-theoretic loss functions for interval estimation have balanced these two criteria: interval size and interval coverage. Restricting the set of decision rules as we have adds upfront difficulty in determining such a set, but enables us to obtain guaranteed coverage and avoids the downstream difficulty of maintaining two optimality criteria in the loss function.

\begin{table}[t]
    \centering
    \caption{Summary of the properties of the intervals considered or mentioned herein. The SSB intervals are the only simultaneous intervals considered, the rest of the methods are designed to work with one functional at a time. ``Empirical Coverage'' indicates if the method covers at the desired level in simulations. ``Provable Coverage'' indicates the existence of a mathematical proof guaranteeing the method's coverage. ``Physical Constraints'' indicates the method's ability to incorporate physical knowledge in the form of affine constraints into the interval construction. ``Rank-Deficient Model'' indicates if the method can be used with a column-rank-deficient linear model, helping mitigate systematic uncertainty due to the binning. \checkmark\textsuperscript{*} indicates that the coverage depends upon sufficiently small systematic error in the linear model used to construct the interval. We assume here that the systematic error can be made sufficiently small for those methods that can handle a rank-deficient matrix. \checkmark\textsuperscript{**} indicates that the method makes use of the constraints but the final intervals do not necessarily respect the physical constraints.
    }
    \label{table:comparison}
    \bigskip
    \resizebox{\textwidth}{!}{\begin{tabular}{c c  c  c  c  c  c}
         \hline
         Interval Type & Coverage Design & Interval Width & Empirical Coverage & Provable Coverage & Physical Constraints & Rank-Deficient Model \\
         \hline \hline
         Tikhonov/Regularized & One-at-a-time & Narrow & \xmark & \xmark & \xmark & \checkmark \\
         \hline
         Least-Squares & One-at-a-time & Medium & \checkmark\textsuperscript{*} & \checkmark & \xmark & \xmark \\
         OSB & One-at-a-time & Medium & \checkmark & \textbf{?} & \checkmark & \checkmark \\
         PO & One-at-a-time & Medium & \checkmark & \checkmark & \checkmark\textsuperscript{**} & \checkmark \\
         \hline
         SSB & Simultaneous & Wide & \checkmark & \checkmark & \checkmark & \checkmark \\
         Minimax & One-at-a-time & Wide & \checkmark\textsuperscript{*} & \checkmark & \checkmark & \xmark  \\
        \hline
    \end{tabular}}
\end{table}
More generally, we have shown with the decision-theoretic approach how one can effectively use ``prior'' information available in the form of a prior distribution
to construct confidence intervals that still maintain frequentist coverage.
If the prior is indeed correct, then the
method would be optimal (in a decision-theoretic sense), but even if the prior
was wrong, the method still provides correct coverage at the cost of the
width of the interval. In a sense, the method is ``tuned" using
the prior information, while ensuring frequentist ``validity."
Broadly, our work falls under an umbrella of work \cite{grunwald_2018,grunwald_deheide_koolen_2020}, among others, that
unites the Bayesian and frequentist perspectives, providing direct ways of accommodating prior information
while keeping frequentist coverage guarantees.

Additionally, the decision-theoretic framework elucidates a practical computational advantage of the PO intervals over the OSB and SSB intervals. Namely, since the optimal decision rule is computed based solely on the functional of interest $\bm{h}$, the forward operator $\bm{K}$, and a given prior, the convex program~\eqref{int:po} only has to be solved once to find a sequence of intervals for a sequence of observations. Since the interval computation requires only vector-vector multiplication given an arbitrary decision rule, computing a sequence of intervals requires relatively little computation. By contrast, no such ``pre-computing" can be done for the OSB and SSB intervals, meaning that for each new data vector, the full end point convex programs must be solved to compute each interval. The ``pre-computed" nature of the PO intervals provides therefore an advantage in applications with a fixed problem setting and a stream of data for which confidence intervals are desired.

The above results provide strong evidence that the OSB and PO intervals work well in the wide-bins-via-narrow-bins unfolding paradigm, but we imagine several immediate next steps to build upon these results. One, as stated in Section~\ref{sec:osb_description}, despite having shown good empirical coverage results in a variety of contexts, the OSB intervals do not yet have a mathematical guarantee that they cover for arbitrary functionals. This is an important future direction as a rigorous proof of the coverage guarantee would provide a solid basis for the use of these intervals in scientific applications. Two, there are a variety of possible configurations we can explore in the decision-theoretic setup for the PO intervals. In particular, we could broaden the class of decision rules beyond affine rules to non-linear rules, and we could explore different loss functions beyond the interval width that take higher-order information into account. It would be interesting to explore how these different decision rules and loss functions affect the interval widths and their sensitivity to prior choice. Third, the results presented herein are based upon simulated data, so applying this method to real data would be a cogent next step. Clearly, applications to real data could not directly assess interval coverage, but they would provide more realistic comparisons of unfolded interval widths between the OSB and PO intervals and other commonly used methods. Fourth, while we have provided a solution to the systematic error stemming from the wide-bin bias, our approach still assumes knowledge of the smearing kernel $k$. In reality, since this kernel is also uncertain, a next step would be to develop methods for incorporating this uncertainty into the final uncertainty quantification. Finally, there is an important middle-ground between the one-at-a-time nature of the OSB and PO intervals and the simultaneous SSB intervals, which by construction provide coverage for an infinite set of functionals (i.e., the set of all linear functionals). Namely, finding $n$-at-a-time intervals or other $n$-variate confidence sets with a coverage guarantee that holds simultaneously for a finite number of $n$ functionals would be useful for scientific inference, as such uncertainty quantification might be more fitting for answering questions such as how well unfolded results comport with theory predictions. The decision-theoretic framework seems like a useful starting point for this extension.

\section*{Acknowledgments}
We are grateful to the members of the CMS Statistics Committee, the CMU Statistical Methods for the Physical Sciences (STAMPS) Research Group and the NSF AI Planning Institute for Data-Driven Discovery in Physics, as well as Amy Braverman, Jonathan Hobbs and Peyman Tavallali, for insightful discussions on inverse problems, uncertainty quantification and unfolding throughout this work.

\clearpage

\bibliographystyle{siamplain}
\bibliography{refs}



\section*{Supplementary material}

# SUPPLEMENTARY MATERIALS: Uncertainty quantification for wide-bin unfolding: one-at-a-time strict bounds and prior-optimized confidence intervals[*]

Michael Stanley[†], Pratik Patil[‡], and Mikael Kuusela[§]

**SM1. Computing the Adversarial Ansatz.** The ansatz $f^{\text{MC}}$ used to compute the matrix $\boldsymbol{K}$ (per Eq. (2.10)) is a source of systematic error. As seen in Section 4, the least-squares intervals in the wide-bin setting are unable to overcome the minor misspecification of the GMM Ansatz, but this shortcoming can be addressed by increasing the number of true bins and then aggregating those bins to the same wide-bin resolution. The OSB and PO intervals are also able to handle this misspecification the same way. But, to motivate the need to handle rank-deficient linear models, we construct the Adversarial Ansatz as an ansatz that provides enough systematic error to depress the least-squares and OSB intervals' empirical coverages significantly below their nominal levels, thus warranting the use of more true bins. We call this constructed ansatz "adversarial" since it is made for the explicit purpose of breaking the empirical coverage of these methods. We construct such an ansatz by leveraging the observations that this problem is ill-posed (and hence inversions are sensitive to noise) and that least-squares estimators have high variance in the absence of regularization.

The brute-force procedure for generating the Adversarial Ansatz is described in Algorithm SM1.1 below. On a high level, Algorithm SM1.1 generates an ensemble of potential ansatz functions, estimates the bin-wise coverage of the least-squares intervals for each ansatz, and then identifies the ensemble element with the lowest minimum estimated bin-wise coverage. The resulting ansatz $f^{\text{MC}}$ from this procedure is shown in Figure 1a. Each potential ansatz function is constructed by generating a realization of data in the smeared space, estimating the true bin counts via non-negatively constrained least-squares estimation, followed by a cubic spline interpolation of the least-squares estimator. This constrained least-squares estimator is very sensitive to noise and hence looks nothing like the true bin means (see the right panel of Figure SM1). A key feature of this construction is that it creates an oscillatory ansatz (see Figure 1a) that when mapped back into the smeared space, fits the data almost exactly (see the left panels of Figures SM1 and 1b). Hence this ansatz could not be ruled out based on the smeared observations. Because of its oscillatory nature, this ansatz creates more systematic error than the Misspecified GMM Ansatz.

[†]Department of Statistics and Data Science, Carnegie Mellon University, Pittsburgh, PA 15213 (mcstanle@andrew.cmu.edu).
[‡]Department of Statistics and Data Science and Machine Learning Department, Carnegie Mellon University, Pittsburgh, PA 15213 (pratik@cmu.edu).
[§]Department of Statistics and Data Science and NSF AI Planning Institute for Data-Driven Discovery in Physics, Carnegie Mellon University, Pittsburgh, PA 15213 (mkuusela@andrew.cmu.edu).

---

**Algorithm SM1.1** Brute-Force Construction of the Adversarial Ansatz

**Inputs**:
- $N_c \in \mathbb{N}$: Number of samples for estimating bin-wise coverage.
- $N_a \in \mathbb{N}$: Number of ansatz functions to compute.
- $\boldsymbol{K} \in \mathbb{R}^{40 \times 40}$: The true smearing matrix.
- $\boldsymbol{\lambda} \in \mathbb{R}^{40}$: True bin means.
- $\{\boldsymbol{h}_i\}_{i=1}^{10}$: A sequence of linear functionals, each aggregating four adjacent bins into one wide bin.

**Output**:
- $f^{\text{MC}}$: Adversarial ansatz.

**Procedure**:
1. Sample $\mathbf{y}_1, \ldots, \mathbf{y}_{N_c} \sim \text{Poisson}(\boldsymbol{K}\boldsymbol{\lambda})$.
2. Let $C$ denote an array of length $N_a$ that will store the minimum estimated bin-wise coverage for each generated ansatz. For $i = 1, \ldots, N_a$:
   (a) Generate data from the true distribution: $\mathbf{y}^i \sim \text{Poisson}(\boldsymbol{K}\boldsymbol{\lambda})$.
   (b) Find the non-negatively constrained least-squares estimator:
   $$\widehat{\boldsymbol{\lambda}}^i = \operatorname*{argmin}_{\boldsymbol{\lambda} \geq \mathbf{0}} \|\mathbf{y}^i - \boldsymbol{K}\boldsymbol{\lambda}\|_2^2.$$
   (c) Interpolate $\widehat{\boldsymbol{\lambda}}^i$ using a cubic spline, defining an ansatz function $f_i^{\text{MC}} : \mathbb{R} \to \mathbb{R}$.
   (d) Compute a smearing matrix $\boldsymbol{K}_i^{\text{MC}}$ using Eq. (2.10) and the previously computed ansatz function $f_i^{\text{MC}}$.
   (e) For each sample $\mathbf{y}_j$ and each function $\boldsymbol{h}_k$ for $j \in [N_c]$ and $k \in [10]$, compute the least-squares interval using Eq. (3.19), and estimate the bin-wise coverage using Eq. (4.5), obtaining a sequence of estimated bin-wise coverages $\{\gamma_l\}_{l=1}^{10}$.
   (f) Find the minimum estimated bin-wise coverage by $\gamma_i^{\min} = \min_{l \in [10]} \gamma_l$, and set $C[i] \leftarrow \gamma_i^{\min}$.
3. Identify the generated ansatz with the lowest estimated bin-wise coverage:
$$i^* = \operatorname*{argmin}_{i \in [N_a]} C[i].$$
4. Return the adversarial ansatz $f_{i^*}^{\text{MC}}$.

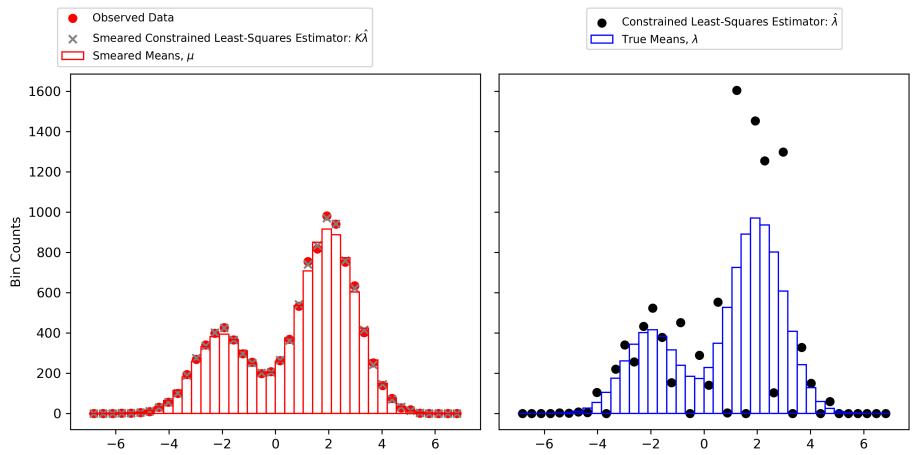

Figure SM1: As shown on the **left** panel, the constrained least-squares estimator ($\widehat{\boldsymbol{\lambda}}$) used to create the Adversarial Ansatz chosen via the minimum bin-wise estimated coverage criterion closely fits the data when mapped back into the smeared space via $\boldsymbol{K}\widehat{\boldsymbol{\lambda}}$, but clearly does not match the true bin means as shown on the **right**.

## SM2. Additional Supporting Figures.

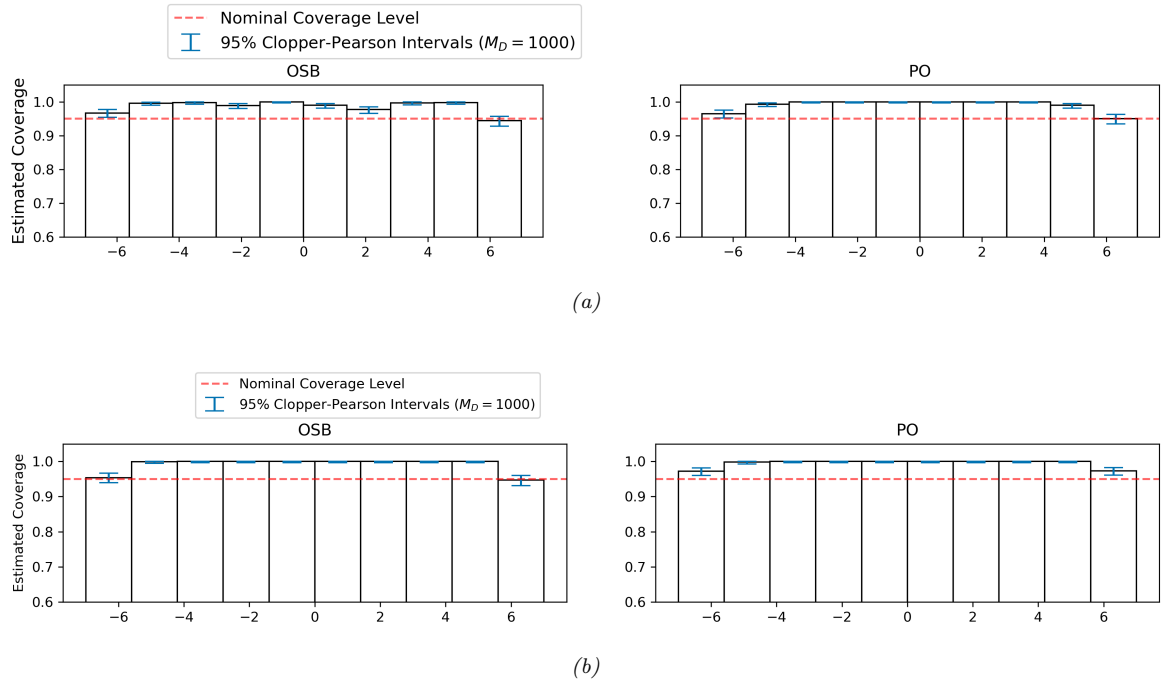

Figure SM2: *Fig. SM2a: Coverage of (from left to right) OSB and PO 95% interval coverage using a flat prior mean constructed from the mean of the true bin expectations. Both methods construct intervals with at least nominal coverage. Fig. SM2b: Coverage of (from left to right) OSB and PO 95% intervals using the $40 \times 80$ smearing matrix constructed with the Adversarial Ansatz. Using a finer true-space binning for the inversion ameliorates the coverage issues displayed in Fig. 5. The least-squares intervals are not applicable in this scenario because of the rank-deficient smearing matrix.*

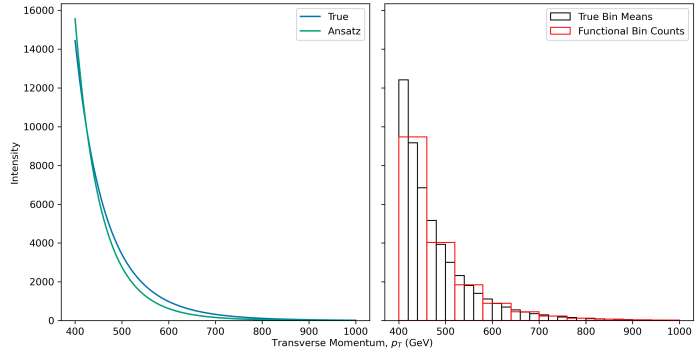

Figure SM3: **(Left)** True and ansatz inclusive jet transverse momentum spectrum intensities. **(Middle)** Scaled (by bin width) expected bin counts for the fine binning (black) and wide binning (red). **(Right)** The middle image but on a logarithmic scale to more clearly see the right-most bins. All these figures show the spectrum in the true space before smearing.

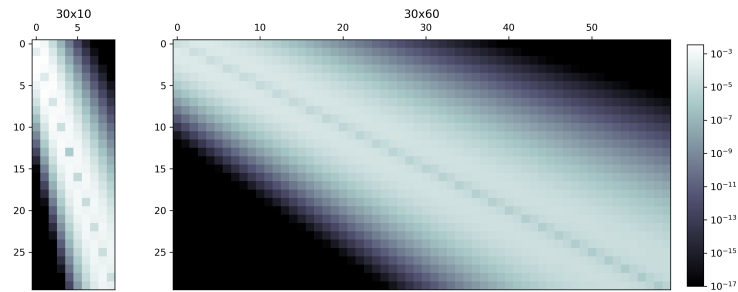

Figure SM4: $|K_{ij} - K_{ij}^{MC}|$ for all rows $i$ and columns $j$ for the **(left)** $30 \times 10$ and **(right)** $30 \times 60$ matrices. The misspecification in the $30 \times 10$ case is $\mathcal{O}(10^2)$ times larger than that of the $30 \times 60$ case.

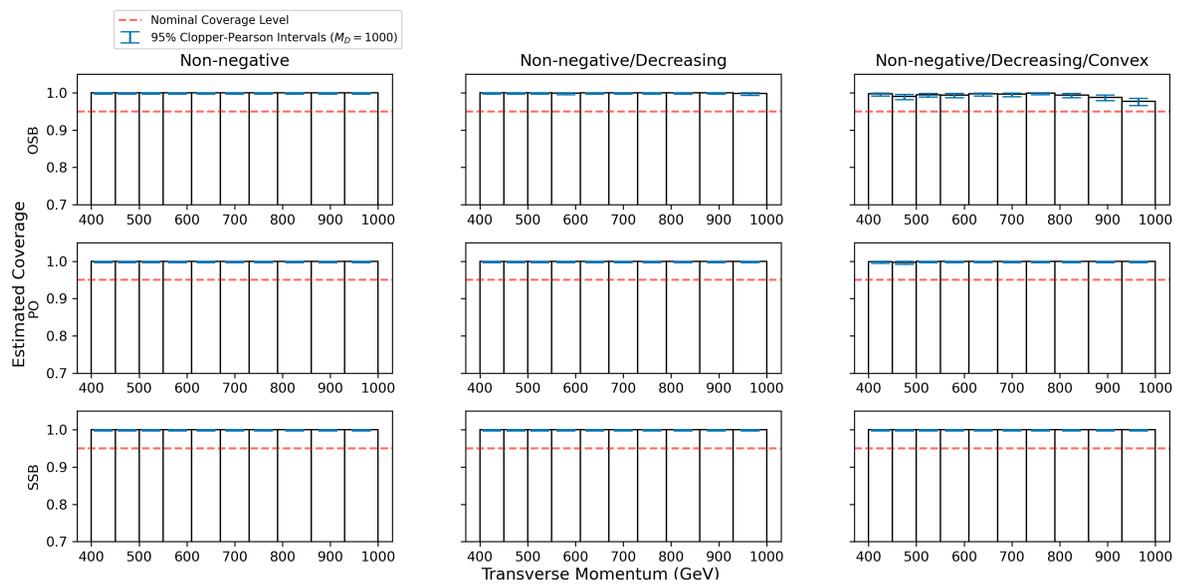

Figure SM5: *Coverage across interval constructions, constraint configurations, and aggregated bins for the steeply falling spectrum. Estimated coverage shows all combinations achieving at least 95% nominal coverage.*



\end{document}